\pdfoutput=1

\documentclass[12pt]{iopart}
\usepackage{iopams}
\usepackage{setstack}
\usepackage{graphicx,epsfig}

\newcommand{\cP}{\ensuremath{\mathcal{P}}}
\newcommand{\cT}{\ensuremath{\mathcal{T}}}
\newcommand{\cPT}{\ensuremath{\mathcal{PT}}}

\newcommand{\vep}{\varepsilon}

\begin{document}
\title[$\cPT$-symmetric potentials having continuous spectra]
{$\cPT$-symmetric potentials having continuous spectra}

\author[Wen and Bender]{Zichao~Wen${}^\ast$ and Carl~M~Bender${}^\dag$}

\address{${}^\ast$Department of Obstetrics and Gynecology, Washington University
School of Medicine, St. Louis, Missouri 63110, USA\\
{\footnotesize{\tt email: zwen@wustl.edu}}}

\address{${}^\dag$Department of Physics, Washington University, St. Louis,
MO 63130, USA \\{\footnotesize{\tt email: cmb@wustl.edu}}}

\date{today}

\begin{abstract}
One-dimensional $\cPT$-symmetric quantum-mechanical Hamiltonians having
continuous spectra are studied. The Hamiltonians considered have the form $H=p^2
+V(x)$, where $V(x)$ is odd in $x$, pure imaginary, and vanishes as $|x|\to
\infty$. Five $\cPT$-symmetric potentials are studied: the Scarf-II potential
$V_1(x)=iA_1\,{\rm sech}(x)\tanh(x)$, which decays exponentially for large $|
x|$; the rational potentials $V_2(x)=iA_2\,x/(1+x^4)$ and $V_3(x)=iA_3\,x/(1+
|x|^3)$, which decay algebraically for large $|x|$; the step-function potential
$V_4(x)=iA_4\,{\rm sgn}(x)\theta(2.5-|x|)$, which has compact support;
the regulated Coulomb potential $V_5(x)=iA_5\,x/(1+x^2)$, which decays slowly
as $|x|\to\infty$ and may be viewed as a long-range potential. The real
parameters $A_n$ measure the strengths of these potentials. Numerical techniques
for solving the time-independent Schr\"odinger eigenvalue problems associated
with these potentials reveal that the spectra of the corresponding Hamiltonians
exhibit universal properties. In general, the eigenvalues are partly real and
partly complex. The real eigenvalues form the continuous part of the spectrum
and the complex eigenvalues form the discrete part of the spectrum. The real
eigenvalues range continuously in value from $0$ to $+\infty$. The complex
eigenvalues occur in discrete complex-conjugate pairs and for $V_n(x)$ ($1\leq n
\leq4$) the number of these pairs is finite and increases as the value of the
strength parameter $A_n$ increases. However, for $V_5(x)$ there is an {\it
infinite} sequence of discrete eigenvalues with a limit point at the origin.
This sequence is complex, but it is similar to the Balmer series for the
hydrogen atom because it has inverse-square convergence.
\end{abstract}
\submitto{\JPA}

\section{Introduction}
\label{s1}
A $\cPT$-symmetric quantum theory is defined by a Hamiltonian that is invariant
under combined space reflection (parity) $\cP$ and time reversal $\cT$. An early
class of non-Hermitian $\cPT$-symmetric Hamiltonians that has been studied in
depth is $H=p^2+x^2(ix)^\vep$ ($\vep$ real). These Hamiltonians are $\cPT$
invariant because $x\to-x$ under $\cP$ and $i\to-i$ under $\cT$. It was observed
in Refs.~\cite{R1,R2} that the eigenvalues of this class of Hamiltonians are
real, discrete, and positive for all $\vep\geq0$ and the reality of these
eigenvalues was attributed to the $\cPT$ symmetry of $H$. Subsequently, the
reality of the spectrum was established at a mathematically rigorous level in
Refs.~\cite{R3,R4}.

The eigenvalues of a $\cPT$-symmetric Hamiltonian are either real or come in
complex-conjugate pairs. If the eigenvalue spectrum is entirely real, the $\cPT$
symmetry of the Hamiltonian is said to be {\it unbroken}, but if some of the
eigenvalues are complex, the $\cPT$ symmetry of the Hamiltonian is said to be
{\it broken}. Many studies of model quantum systems whose Hamiltonians are
$\cPT$ invariant have been published (see Refs.~\cite{R5,R11}). $\cPT$-symmetric
Hamiltonians often exhibit a transition from a parametric region of unbroken
$\cPT$ symmetry to a region of broken $\cPT$ symmetry. This $\cPT$ transition
occurs in both the classical and in the quantized versions of a $\cPT$-symmetric
Hamiltonian \cite{R2} and this transition has been observed in numerous
laboratory experiments \cite{R5,R11}.

Many papers on $\cPT$-symmetric Hamiltonians having discrete spectra have been
published, but there have been only very few studies of $\cPT$-symmetric
Hamiltonians having continuous spectra. Therefore, in this paper we consider
Hamiltonians $H=p^2+V(x)$ whose potentials possess continuous spectra. The
potentials $V(x)$ that we discuss here decay to $0$ as $|x|\to\infty$. Thus, it
is not surprising that we find that in general the real part of the spectrum is
continuous and that these eigenvalues range continuously from 0 to $+\infty$.
Since the potential $V(x)$ is {\it odd} under $x\to-x$ and is pure imaginary,
$V(x)$ is $\cPT$ invariant, and hence complex eigenvalues must appear as
complex-conjugate pairs. On the basis of our study we believe that short-range
potentials typically have a finite number of discrete complex eigenvalues and
that long-range potentials have an infinite number of discrete complex
eigenvalues.

To be specific, in this paper we study five one-dimensional Hamiltonians of the
form $H_n=p^2+V_n(x)$, where
\begin{eqnarray}
V_1(x)&=&iA_1\,{\rm sech}(x)\tanh(x),\nonumber\\
V_2(x)&=&iA_2\,x/(1+x^4),\nonumber\\
V_3(x)&=&iA_3\,x/(1+|x|^3),\nonumber\\
V_4(x)&=&iA_4\,{\rm sgn}(x)\theta(2.5-|x|),\nonumber\\
V_5(x)&=&iA_5\,x/(1+x^2),
\label{e1}
\end{eqnarray}
and the strength parameters $A_n$ are real. In all cases $V_n(x)$ is odd in $x$
and is pure imaginary, so the $H_n$ are all $\cPT$ symmetric. Furthermore, these
potentials all vanish as $|x|\to\infty$. In all these cases the Hamiltonians $H_n$
have real eigenvalues that range continuously from $0$ to $+\infty$. However,
the universal property of the five Hamiltonians studied here is that in addition
to the real continuous part of the spectra, there are complex-conjugate pairs of
discrete eigenvalues for sufficiently large strength parameters $A_n$. Thus, in
all these cases the $\cPT$ symmetry of these Hamiltonians $H_n$ is broken.

The five potentials $V_n(x)$ vanish at different rates for large $|x|$: The
Scarf-II potential $V_1(x)$ decays exponentially for large $|x|$
\cite{R6,R7,R8,R9}, the rational potentials $V_2(x)$ and $V_3(x)$ decay
algebraically like $|x|^{-3}$ and $|x|^{ -2}$ for large $|x|$, and the
step-function potential $V_4(x)$ has compact support \cite{R10}. These are all
{\it short-range} potentials and we find that these potentials confine a {\it
finite} number of discrete complex bound states. The number and size of these
complex eigenvalues increase as the strength parameters $A_n$ increase.

The potential $V_5(x)$ is special; it vanishes slowly like $1/|x|$ for large
$|x|$. Because it vanishes slowly,/and to
it is a {\it long-range} potential like the
Coulomb potential. Even though this potential is bounded, the property that it
is long range allows it to confine {\it infinitely many} discrete complex bound
states. Like the Balmer series for the hydrogen atom, the sequence of
complex-conjugate pairs of eigenvalues converges to a limit point, which happens
to be at $0$, and the $k$th pair of eigenvalues approaches $0$ like $1/k^2$.

In previous studies of $\cPT$-symmetric Hamiltonians it was found that accurate
numerical calculations of real discrete eigenvalues could be done by using the
shooting method \cite{R1}. However, if the discrete eigenvalues are complex, the
shooting method becomes unwieldy. Therefore, alternative techniques based on the
finite-element and variational methods were used. A numerical technique known as
the {\it Arnoldi algorithm} was used in Ref.~\cite{R12}.

Let us summarize the numerical technique used in this paper: For numerical
calculations of eigenvalues one cannot work directly on the infinite $x$ axis,
so one reduces the problem to solving the Schr\"odinger equation on a large but
{\it finite} interval. Consequently, the numerical techniques used to calculate
eigenvalues can only return discrete values and one must determine whether a
given eigenvalue belongs to a discrete or a continuous part of the spectrum. To
distinguish between these two possibilities we examine the associated
eigenfunctions and observe how they satisfy the boundary conditions. As
explained in detail in Ref.~\cite{R12}, the eigenfunctions associated with
discrete eigenvalues are localized in space (like bound states) and decay to 0
smoothly and exponentially as $x$ approaches the boundary points of the
interval. However, the eigenfunctions for eigenvalues that belong to the
continuous part of the spectrum drop abruptly to $0$ at one or both endpoints
of the finite interval.

The technique used here to compute the eigenvalues of $H_n$ is called {\it
Chebyshev spectral collocation}. This technique relies on the properties of
Chebyshev polynomials and Chebyshev series and is explained in detail in
Ref.~\cite{R13}. To calculate the spectra of the Hamiltonians $H_n$ by using
Chebyshev spectral collocation we replace the infinite $x$ axis by the finite
interval $-L\leq x\leq L$. We then decompose the interval $[-L,L]$ into $N$
subintervals bounded by grid points at $x_j$, where $j=0,~1,~2,~3,~\dots,~N$.
These subintervals are not of equal length; rather, the subintervals shorten
as we approach the endpoints at $x=-L$ and $x=L$. To determine the positions of
the grid points, we construct a semicircle of radius $L$ centered at the origin
$x=0$ and divide the circle into equal sectors. We then project onto the $x$
axis. Thus, the grid points are located at $x_j=L\cos(\pi j/N)$. For all
computations done in this paper we take $N=2^{14}-1$. The first and last grid
points lie at $x=\pm L$, but since $N$ is odd there is no grid point at the
origin. We make this choice because the potential $V_4(x)$ is discontinuous at
$x=0$. To find the eigenvalues we impose homogeneous boundary conditions at the
endpoints $\pm L$ and finally let $L$ tend to infinity. In all cases we take
$L=10$, $100$, and $1000$, and we find that the eigenvalues converge rapidly to
their $L=\infty$ values.

This paper is organized very simply. In Secs.~\ref{s2}-\ref{s6} we describe
in turn the spectra of $V_n(x)$ for $n=1$, 2, 3, 4, and 5, and then in
Sec.~\ref{s7} we give some concluding remarks.

\section{Eigenvalues for the Scarf-II potential $V_1(x)$}
\label{s2}
\begin{figure}[b!]
\begin{center}
\includegraphics[scale=0.50]{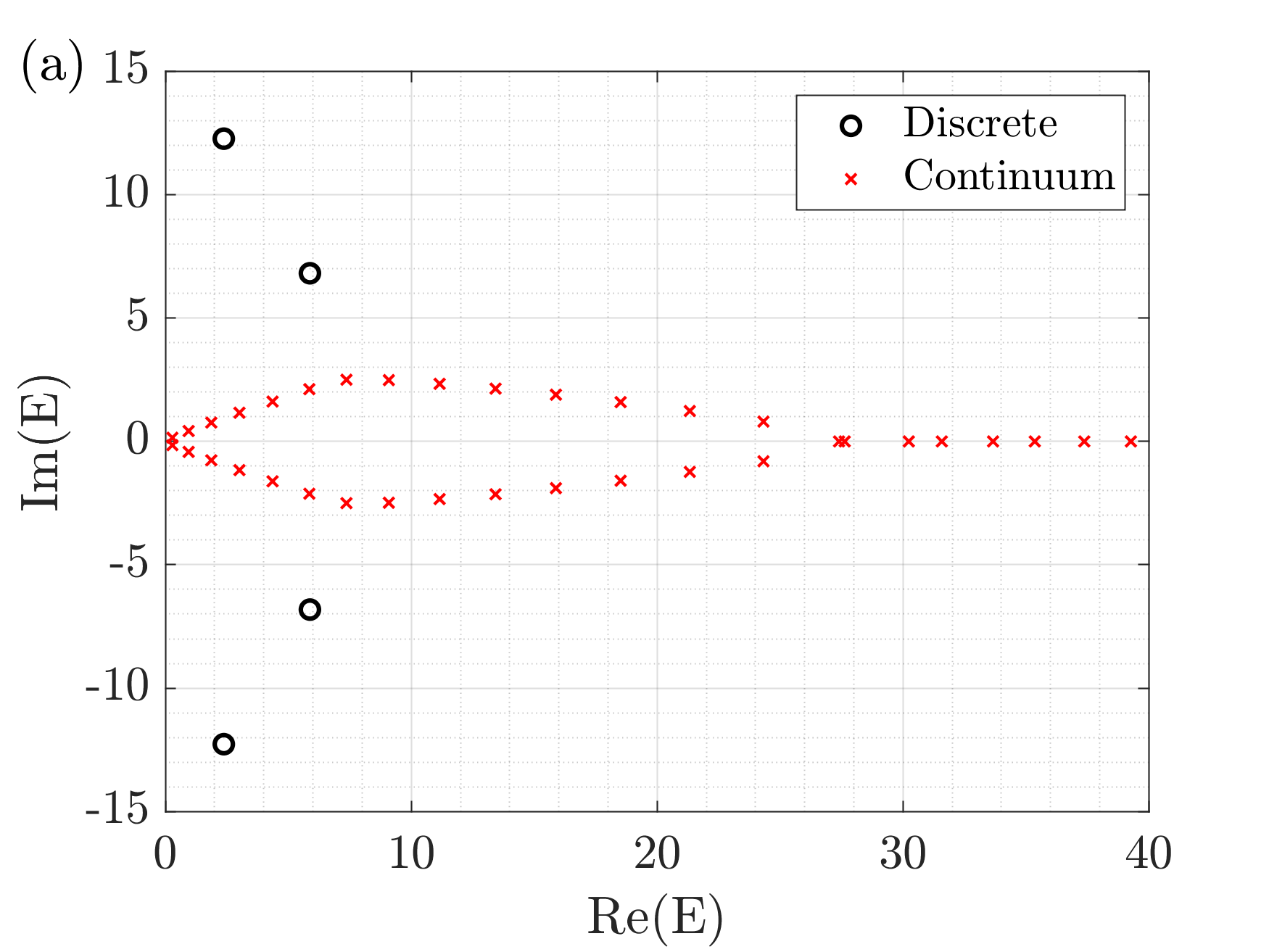}\hspace{-.1cm}
\includegraphics[scale=0.50]{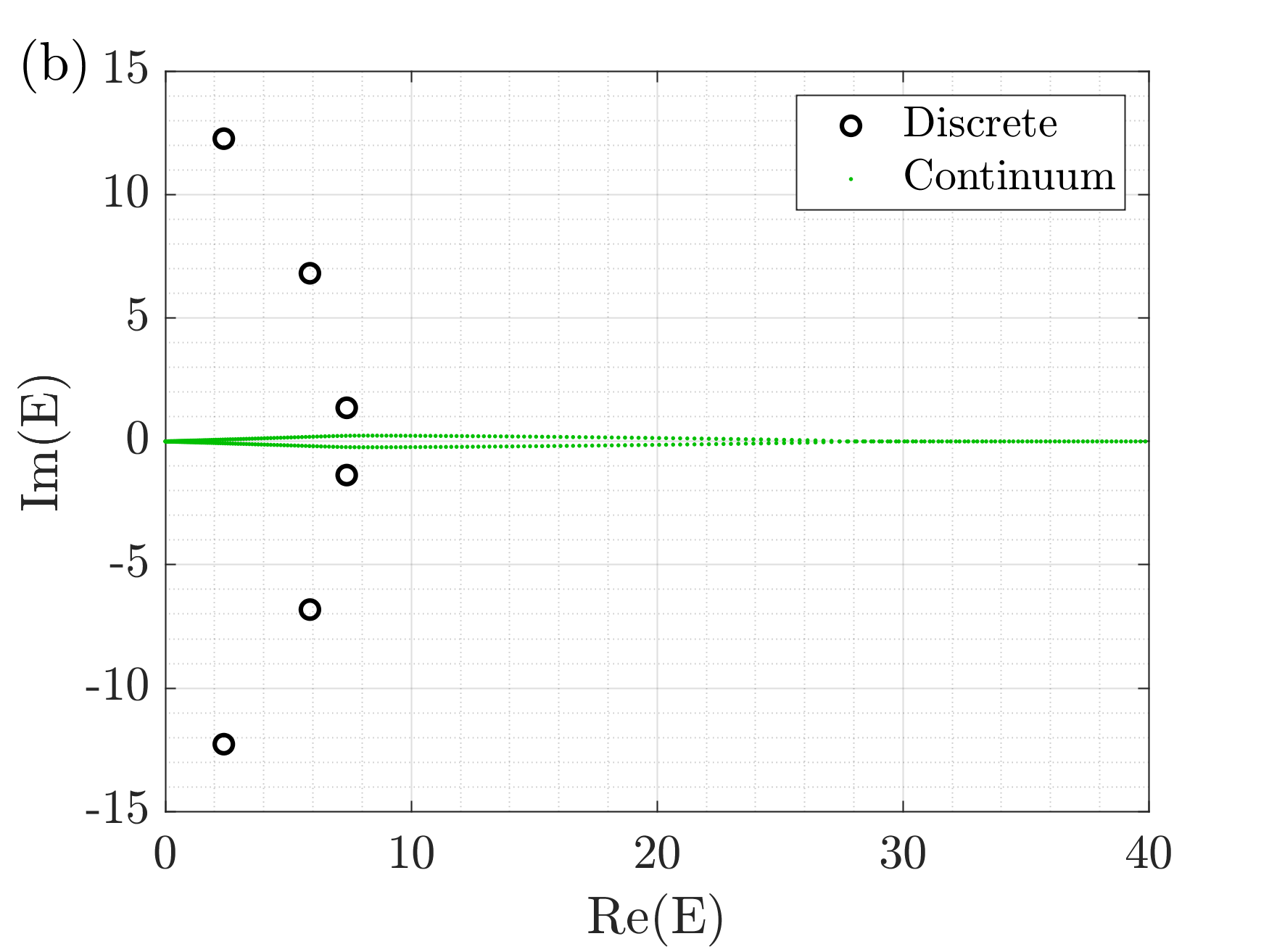}
\end{center}
\caption{[Color online] Eigenvalues in the complex plane for $V_1(x)$ in
(\ref{e1}) with the strength parameter $A_1=30$ for the case $L=10$ in panel
(a) and for $L=100$ in panel (b). Continuum eigenvalues are indicated by red
crosses in panel (a) and by green dots in panel (b). Discrete bound-state
eigenvalues are indicated by black circles. Note that as $L$ increases, the
locations of the bound states stabilize but the continuum eigenvalues all
collapse onto the real axis. As they do so, a new complex-conjugate pair of 
bound states is uncovered. Observe that the continuum eigenvalues come in
complex-conjugate pairs until the real parts of these eigenvalues exceeds about
$28$. Above this critical value the continuum eigenvalues are all real.}
\label{F1}
\end{figure}

\begin{figure}[t!]
\begin{center}
\includegraphics[scale=0.50]{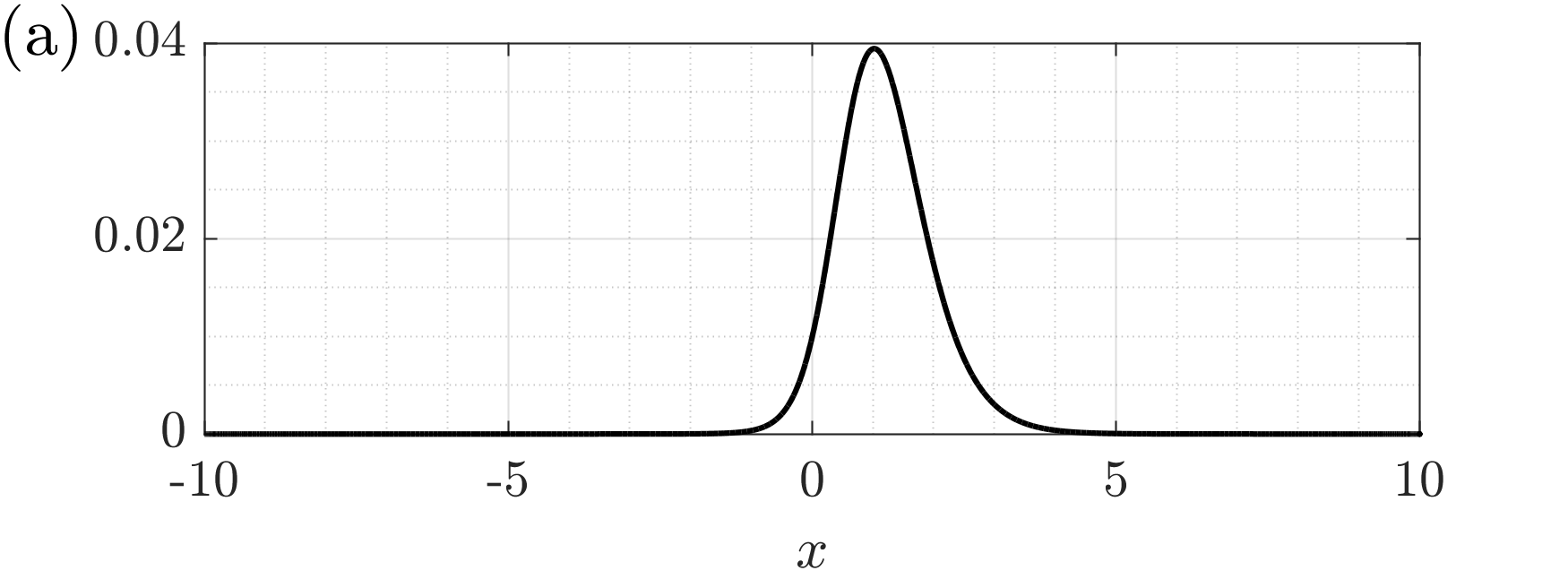}\hspace{-.1cm}
\includegraphics[scale=0.50]{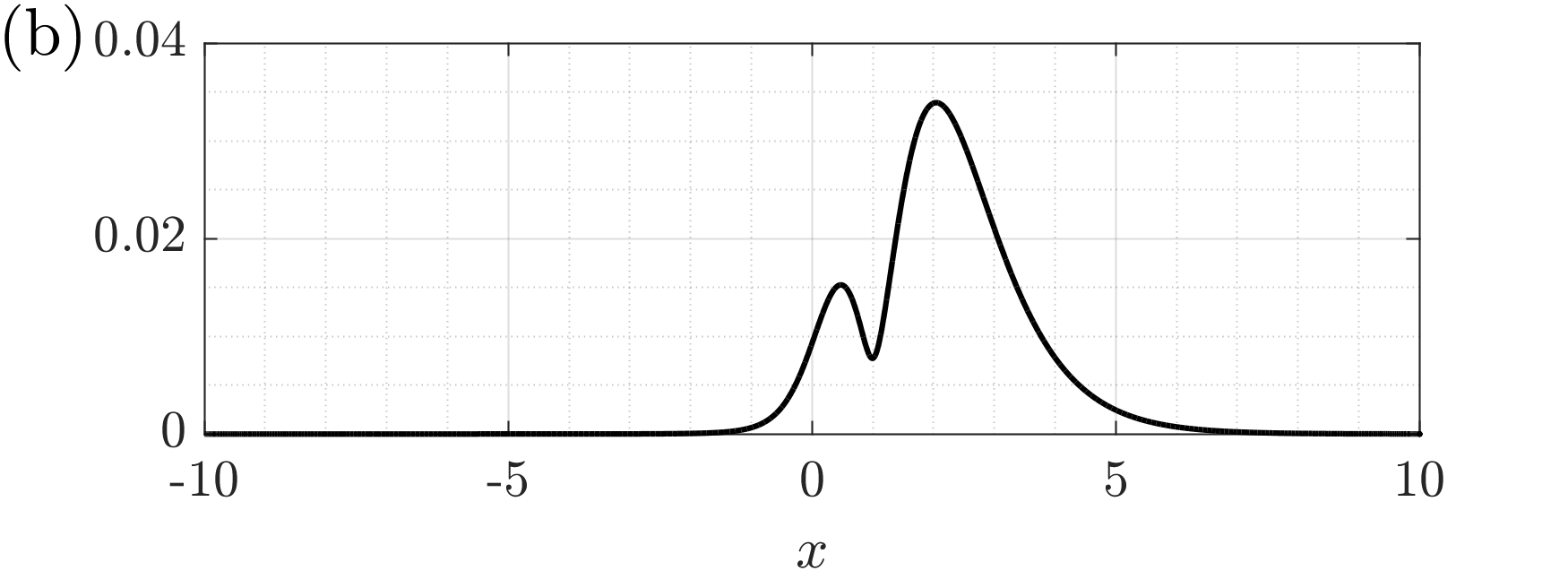}\\
\includegraphics[scale=0.50]{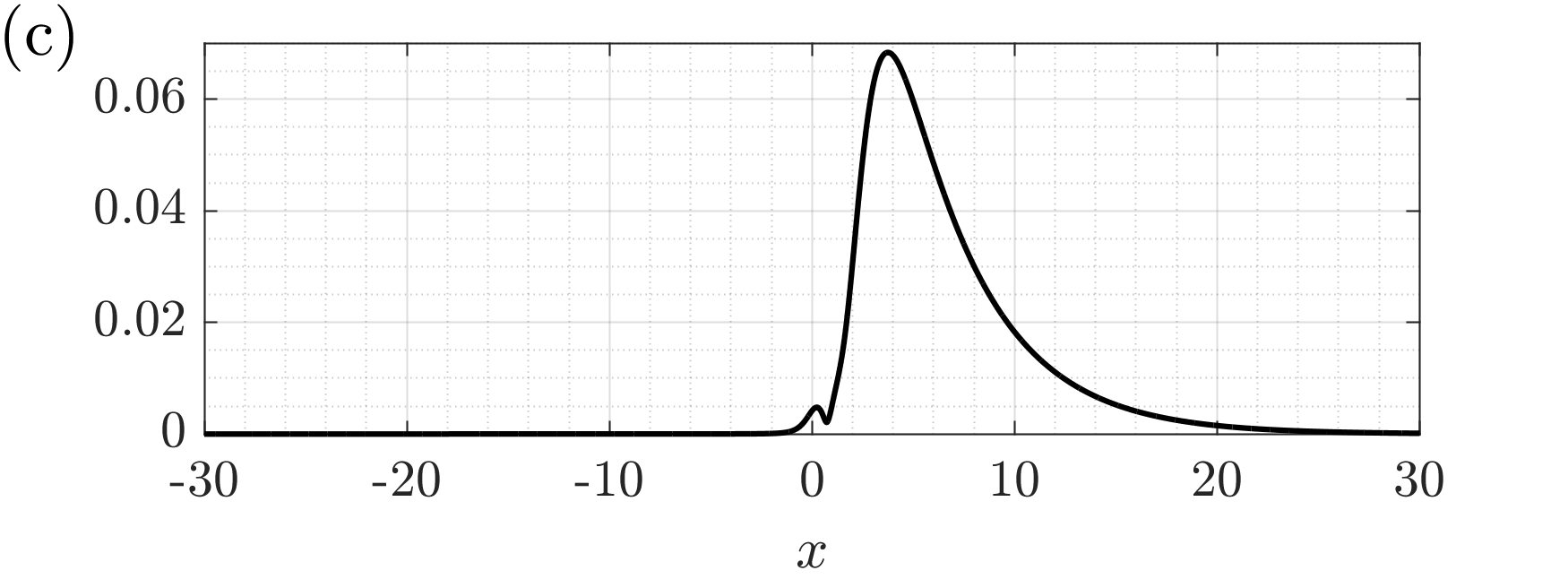}\hspace{-.1cm}
\includegraphics[scale=0.50]{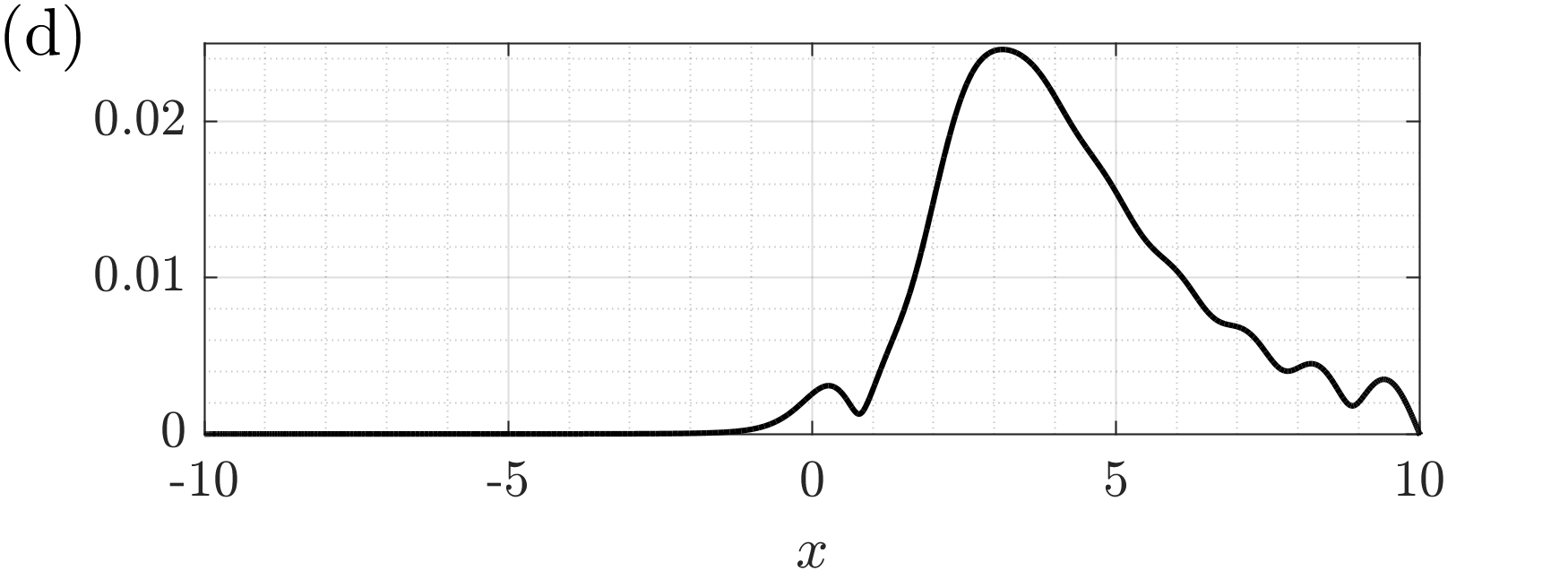}\\
\includegraphics[scale=0.50]{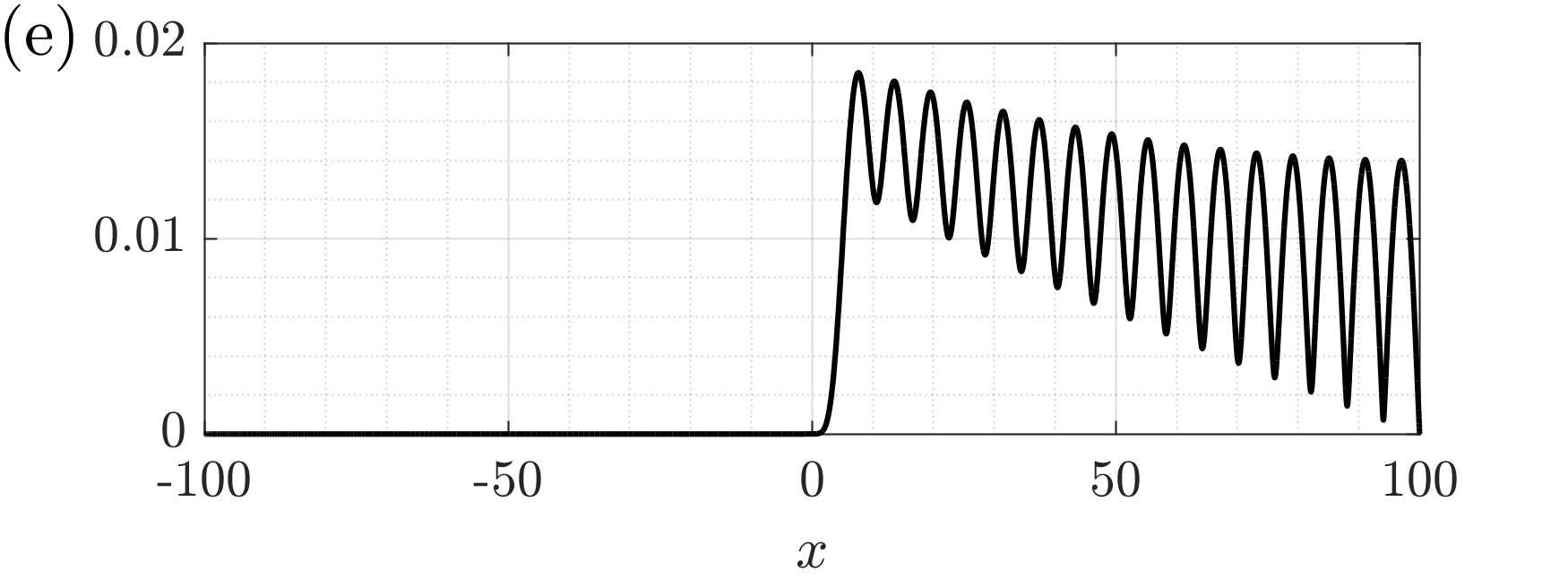}\hspace{-.1cm}
\includegraphics[scale=0.50]{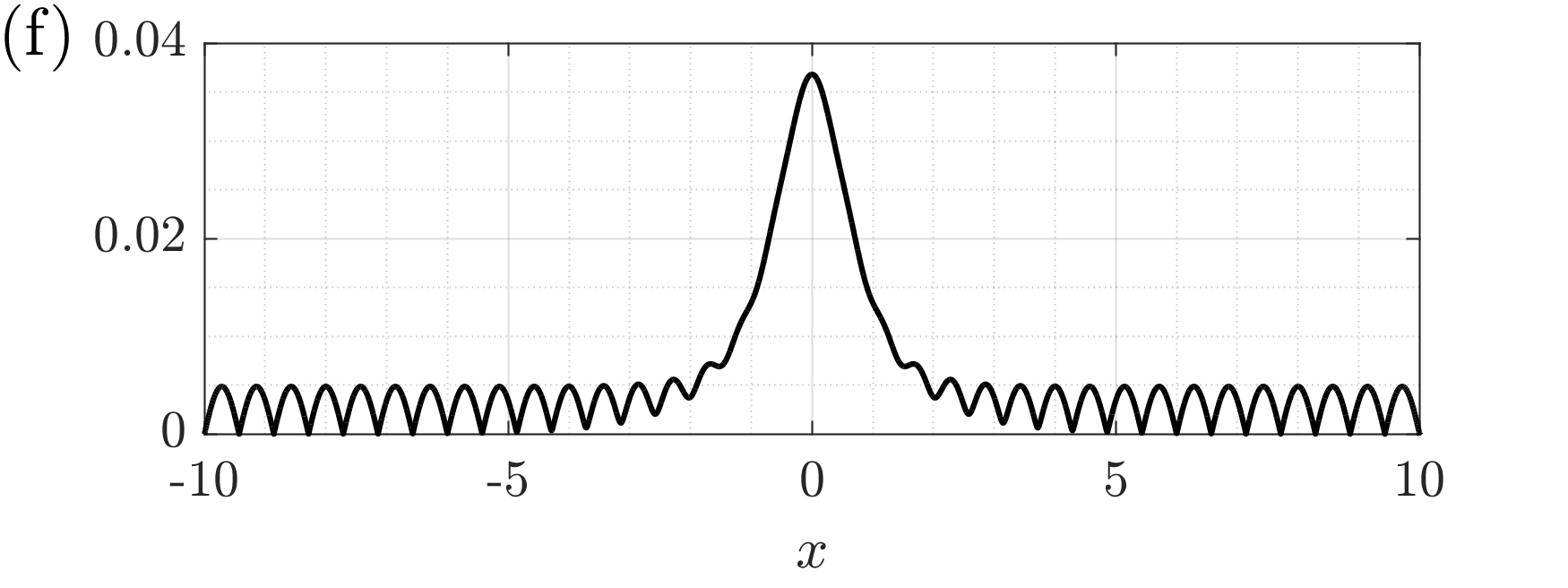}
\end{center}
\caption{[Color online] Plots of the absolute value of the eigenfunctions
associated with some eigenvalues in Fig.~\ref{F1}. Panels (a), (b), and (c)
display the eigenfunctions of the three discrete bound-state eigenvalues in
panel (b) in Fig.~\ref{F1}: Panel (a) shows the eigenfunction for the eigenvalue
$2.374999999702702+12.272301129148877\,i$ for $L=10$; panel (b) shows the
eigenfunction for the eigenvalue $5.875000000021835 +6.817945071620461\,i$ for
$L=10$; panel (c) shows the eigenfunction for the eigenvalue $7.374999997301000+
1.363589013462076\,i$ for $L=100$. The next three plots show the behavior of the
eigenfunctions for some continuum eigenvalues: Panel (d) shows the eigenfunction
for the continuum eigenvalue $7.361943725638523+2.501634415578858\,i$ for $L=
10$; panel (e) shows the eigenfunction for the continuum eigenvalue
$0.277713365597523+0.009073644654185\,i$ for $L=100$; panel (f) shows the
eigenfunction for the continuum eigenvalue $30.234638465149410+0.000000000526705
\,i$ for $L=10$. Bound-state eigenfunctions decay smoothly and exponentially as
$x$ approaches the endpoints but the continuum eigenfunctions abruptly and
sharply drop to $0$ at one or both endpoints.}
\label{F4}
\end{figure}

For the potential $V_1(x)$ in (\ref{e1}) we have chosen $A_1=30$. In
Fig.~\ref{F1} we plot the eigenvalues in the complex plane for $L=10$ in panel
(a) and for $L=100$ in panel (b). We observe two kinds of eigenvalues,
bound-state eigenvalues, which are indicated by circles, and continuum
eigenvalues, which are indicated by crosses in (a) and by dots in (b). As we
will see in Fig.~\ref{F4}, we can distinguish between bound-state and continuum
eigenvalues by examining the corresponding eigenfunctions. This plot of the
absolute values of the eigenfunctions as functions of $x$ shows that the
eigenfunctions for bound-state eigenvalues decay smoothly and exponentially to 0
as $x$ approaches the boundaries at $\pm L$ while the continuum eigenfunctions
abruptly drop to 0 at one or both boundaries. Note that as $L$ increases from
$10$ in (a) to $100$ in (b), the positions of the bound-state eigenvalues
stabilize rapidly but the continuum eigenvalues approach the real axis.

\begin{figure}[t!]
\begin{center}
\includegraphics[scale=0.50]{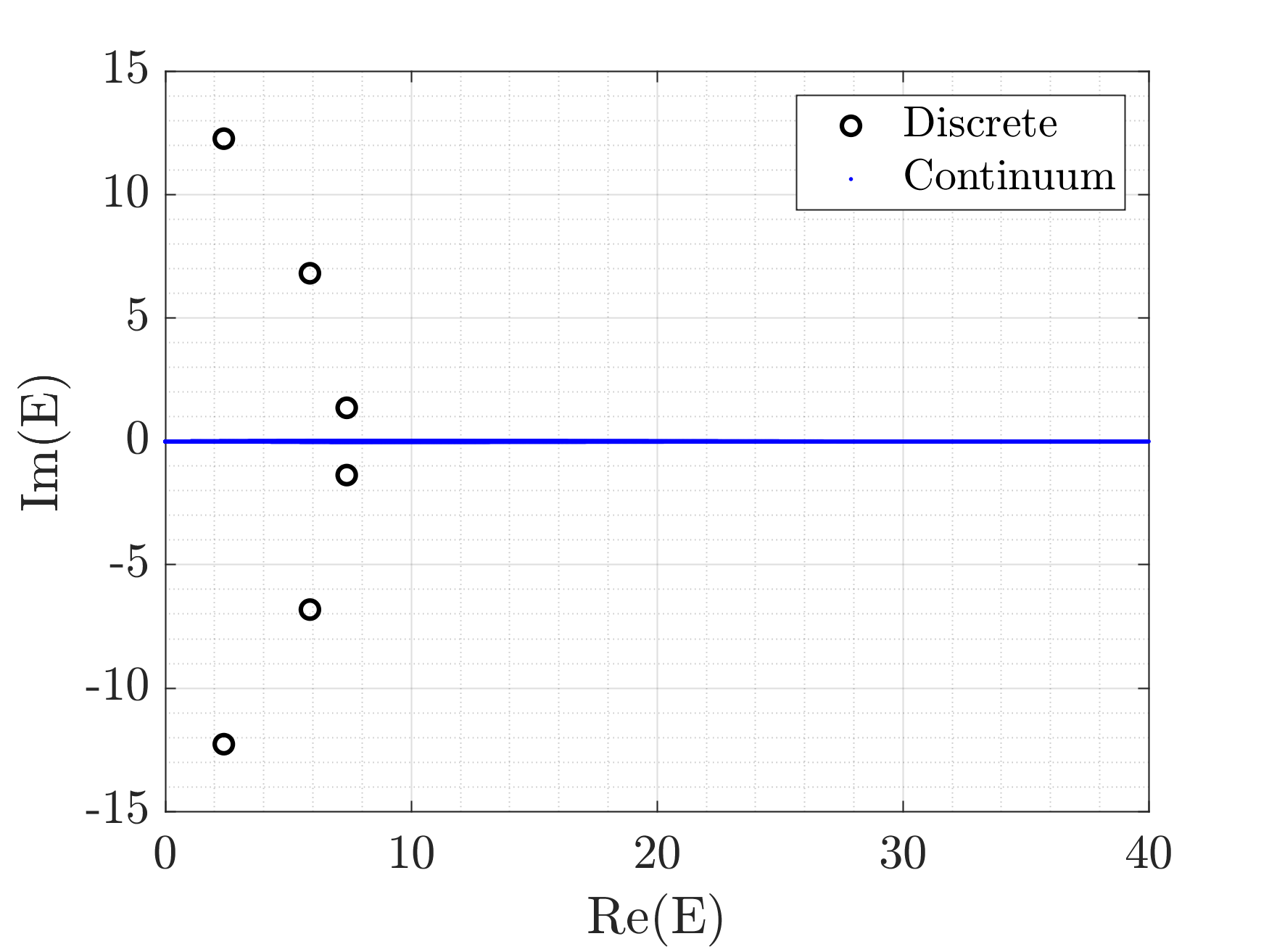}
\end{center}
\caption{[Color online] Eigenvalues in the complex plane for $V_1(x)$ in
(\ref{e1}) with the strength parameter $A_1=30$ for the case $L=1000$. The
discrete bound-state eigenvalues (black circles) have not moved but the
continuum eigenvalues (blue dots) are now very close to the real axis.}
\label{F2}
\end{figure}

We observe two kinds continuum eigenvalues in Fig.~\ref{F1}. Above a critical
value near $28$ the continuum eigenvalues are real, but below this critical
value the continuum eigenvalues come in complex-conjugate pairs that lie
slightly above and below the real axis. These pairs of eigenvalues approach the
real axis as $L$ increases but the position of the critical value near $x=28$
does not change. With increasing $L$ each member of the complex-conjugate pair
of eigenvalues approaches the real axis vertically, but as they reach the real
axis, one member of the pair moves slightly rightward and the other moves
slightly leftward along the real axis, thus doubling the density of points.
(This behavior of the continuum eigenvalues is in exact analogy to the motion of
the roots of the famous Wilkenson polynomial \cite{R14}.) For $L=1000$, the
continuum eigenvalues below $x=28$ are extremely close to the real axis, but the
positions of the complex bound states do not move, as we can see in
Fig.~\ref{F2}.

\begin{figure}[h!]
\begin{center}
\includegraphics[scale=0.50]{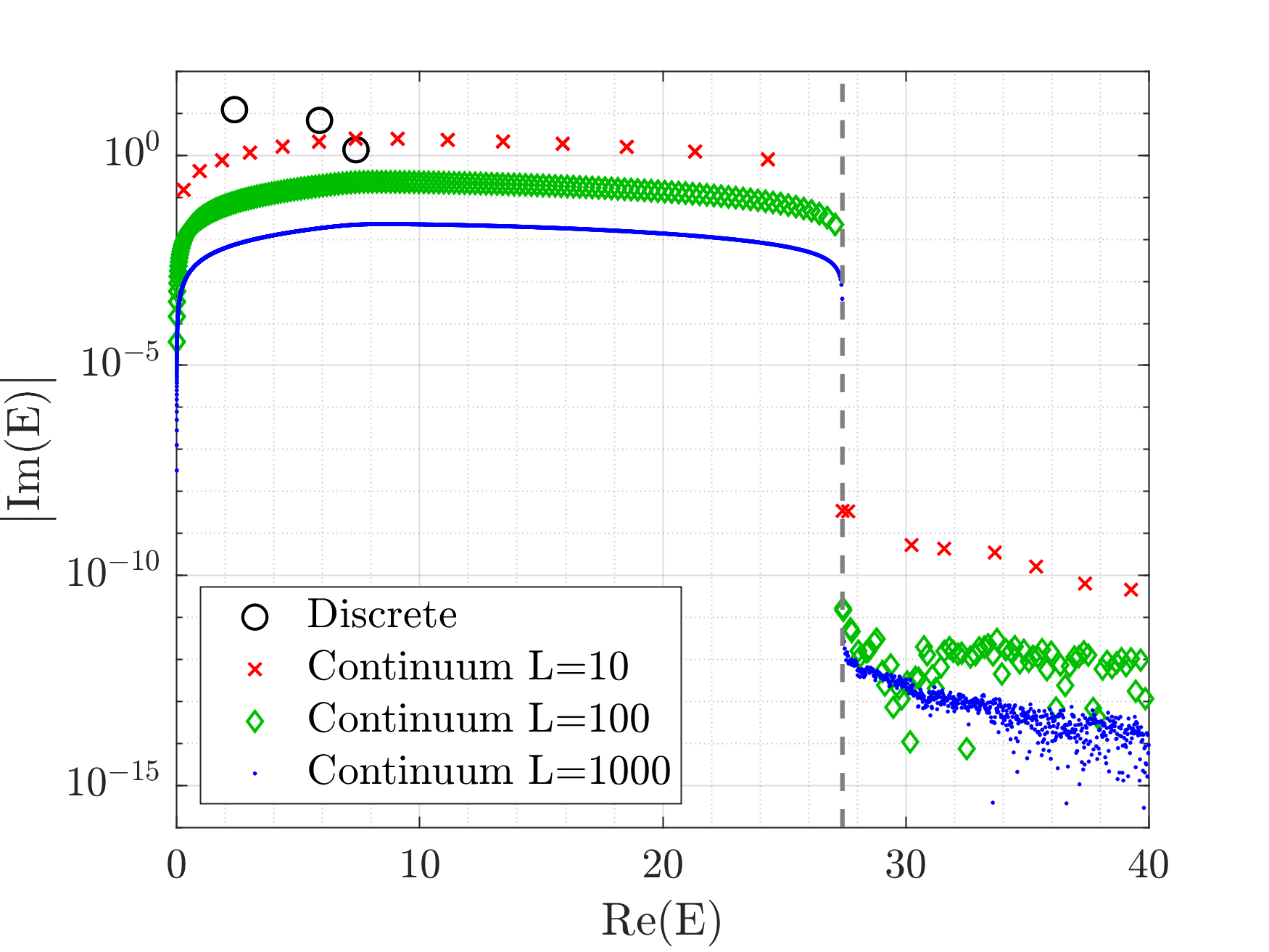}
\end{center}
\caption{[Color online] Logarithmic plot of the eigenvalue data in
Figs.~\ref{F1} and \ref{F2}. This plot presents dramatic evidence that the
transition from slightly complex continuum eigenvalues to exactly real
continuum eigenvalues near $x=28$ is sharp. There is a jump at the dashed line
of about {\it ten orders of magnitude} at this transition point. Observe
that the location of the transition does not change as $L$ is increased from
$10$ to $100$ to $1000$.}
\label{F3}
\end{figure}

Figure~\ref{F3} is a logarithmic plot of the eigenvalues shown in Figs.~\ref{F1}
and \ref{F2}. Observe that the critical point at $x=28$ at which the continuum
eigenvalues jump from being complex-conjugate pairs to real numbers does not
move as $L$ is increased.

To distinguish between discrete and continuum eigenvalues we investigate the
behavior of the associated eigenfunctions. Six possible behaviors of the
eigenfunctions are displayed in Fig.~\ref{F4}. In general, the eigenfunctions of
discrete eigenvalues decay smoothly to 0 at the endpoints of the interval but
the eigenfunctions of continuum eigenvalues abruptly drop to 0 at one or both
endpoints.

\section{Eigenvalue behavior of $V_2(x)$}
\label{s3}
\begin{figure}[t!]
\begin{center}
\includegraphics[scale=0.50]{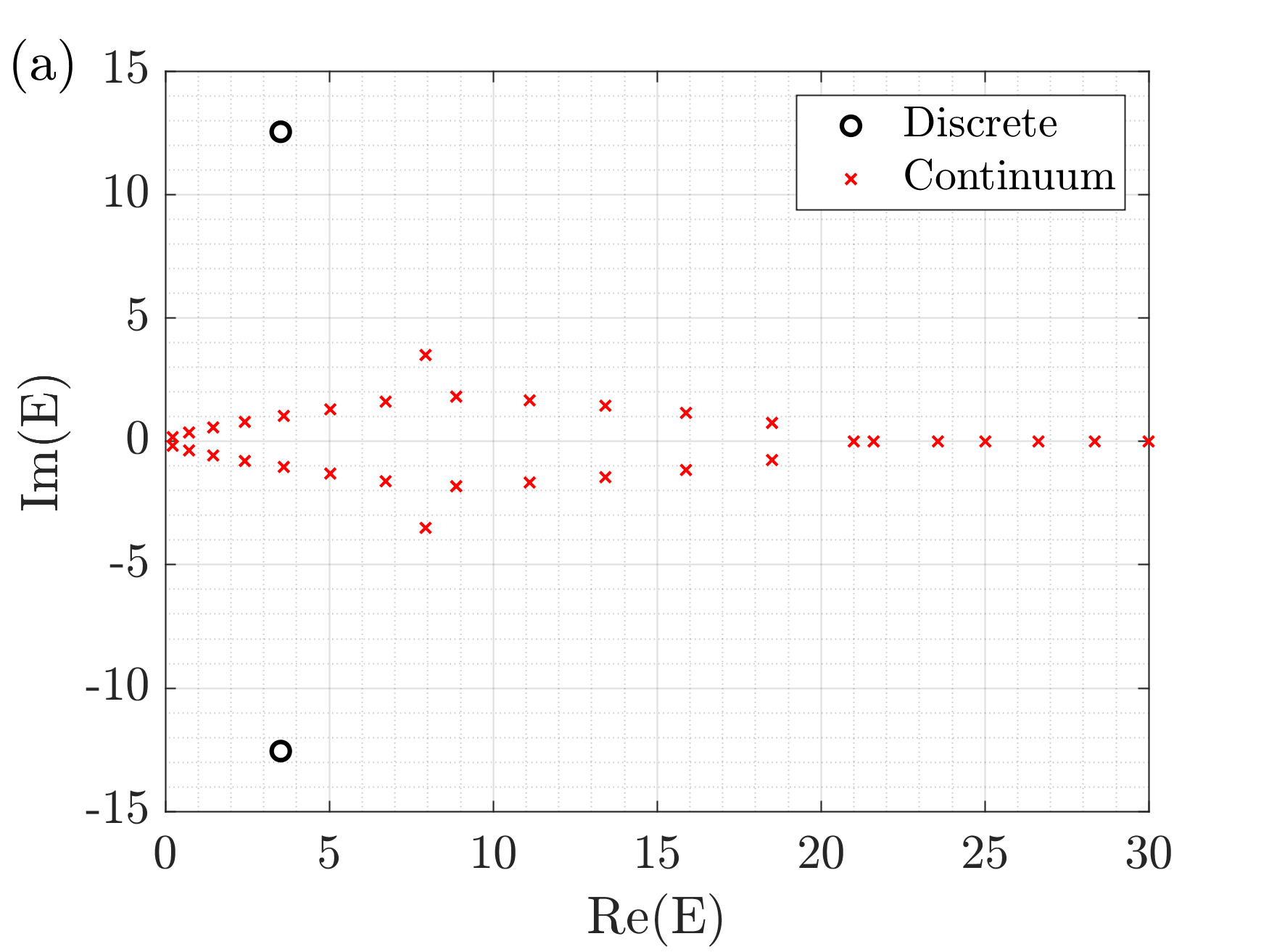}\hspace{-.1cm}
\includegraphics[scale=0.50]{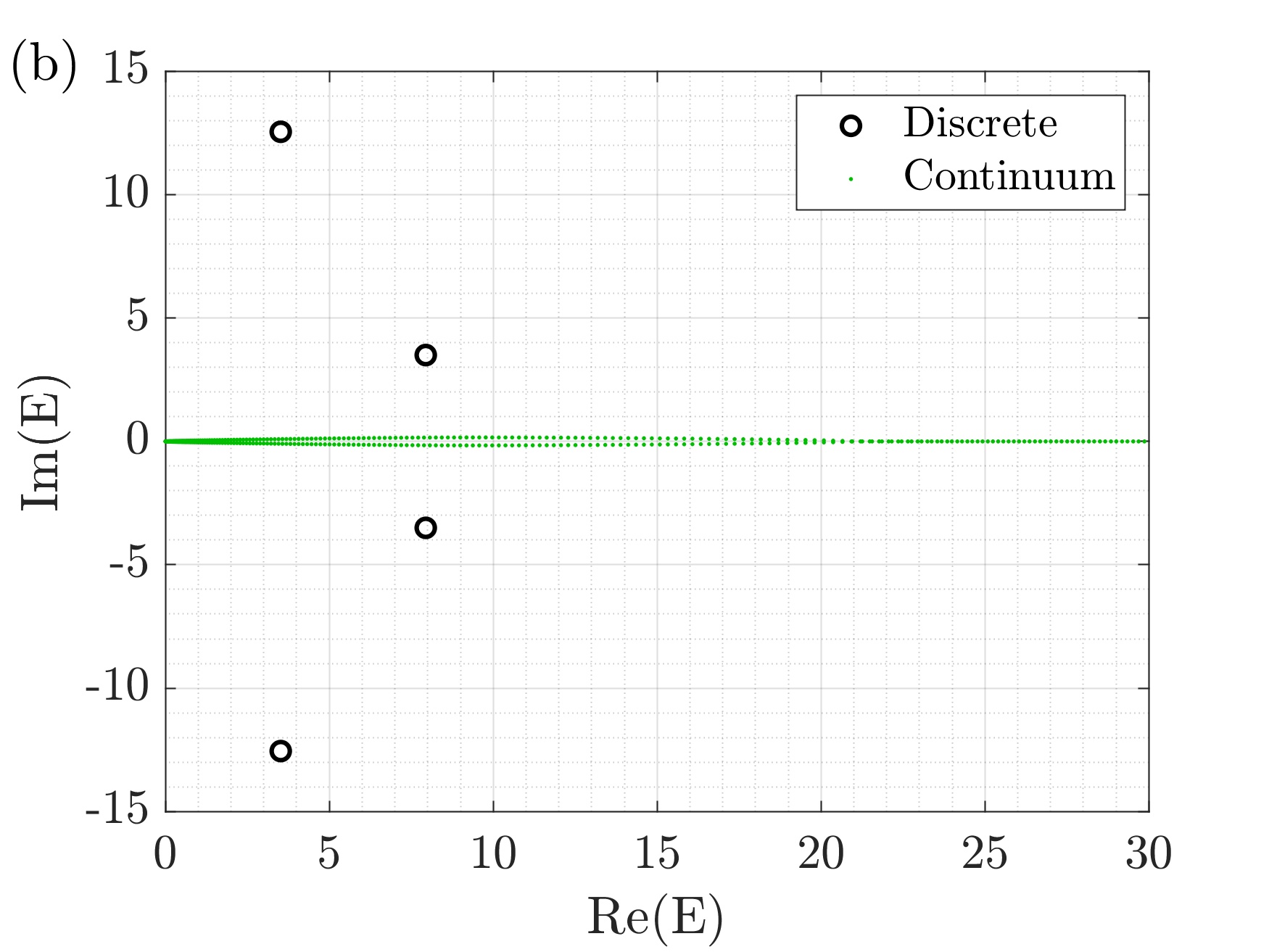}
\end{center}
\caption{[Color online] Energy spectra for $V_2(x)$ with $A_2=30$. Panel (a)
shows the eigenvalues for $L=10$ and panel (b) shows the eigenvalues for
$L=100$. Like the case for the Scarf-II potential $V_1(x)$, the continuum
part of the spectrum is slightly complex until the critical point near $21$,
after which the continuum eigenvalues are real. One pair of bound-state
eigenvalues (black circles) is visible in panel (a) but as $L$ is increased
to $100$ in panel (b), a new pair of bound-state eigenvalues is uncovered.}
\label{F5}
\end{figure}

While the potential $V_1(x)$ decays exponentially for large $|x|$, the
potential $V_2(x)$ decays algebraically like $|x|^{-3}$ for large $|x|$.
Nevertheless, the spectral properties of $V_2(x)$ are strikingly similar to
those of $V_1(x)$. For $V_2(x)$ the analog of Fig.~\ref{F1} is Fig.~\ref{F5}.
Again, we have taken the strength parameter $A_2$ to be $30$ and we observe one
complex-conjugate pair of bound-state eigenvalues for $L=10$ in panel (a) and
two complex-conjugate pairs of bound-state eigenvalues for $L=100$ in panel (b).
The new pair of bound-state eigenvalues is uncovered as the continuum
eigenvalues collapse towards the real axis. 

\begin{figure}[h!]
\begin{center}
\includegraphics[scale=0.50]{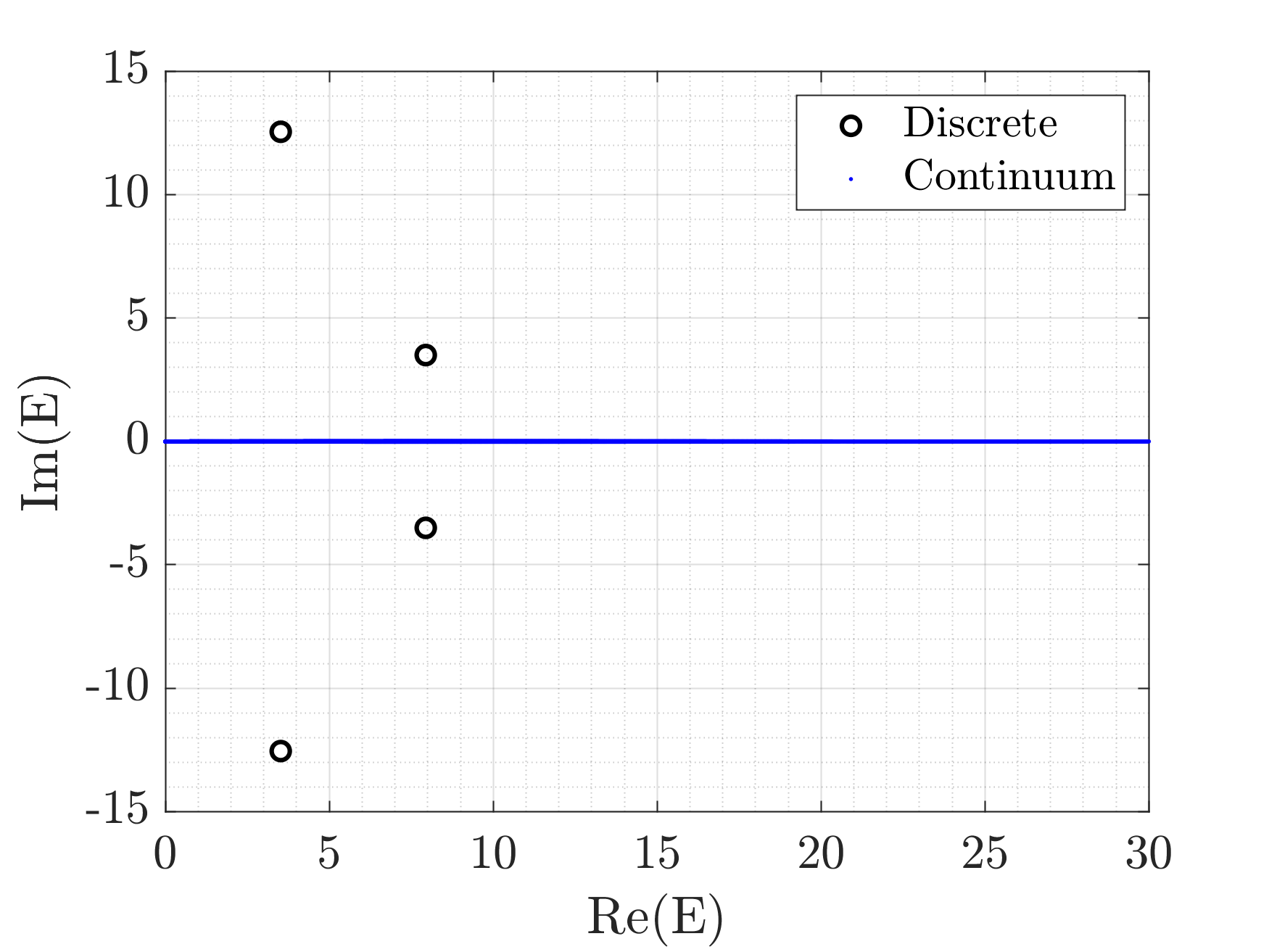}
\end{center}
\caption{[Color online] Energy eigenvalues for the $V_2(x)$ potential for
$A_2=30$. For this calculation $L=1000$. Note that the discrete bound-state 
energies have not changed from their values in Fig.~\ref{F5}(b). However, the 
continuum part of the spectrum has moved closer to the real axis.}
\label{F6}
\end{figure}

If we increase $L$ to $1000$, we observe in Fig.~\ref{F6} that the bound-state
eigenvalues do not move. However, the continuum eigenvalues lie very close to
the real axis.

\begin{figure}[t!]
\begin{center}
\includegraphics[scale=0.50]{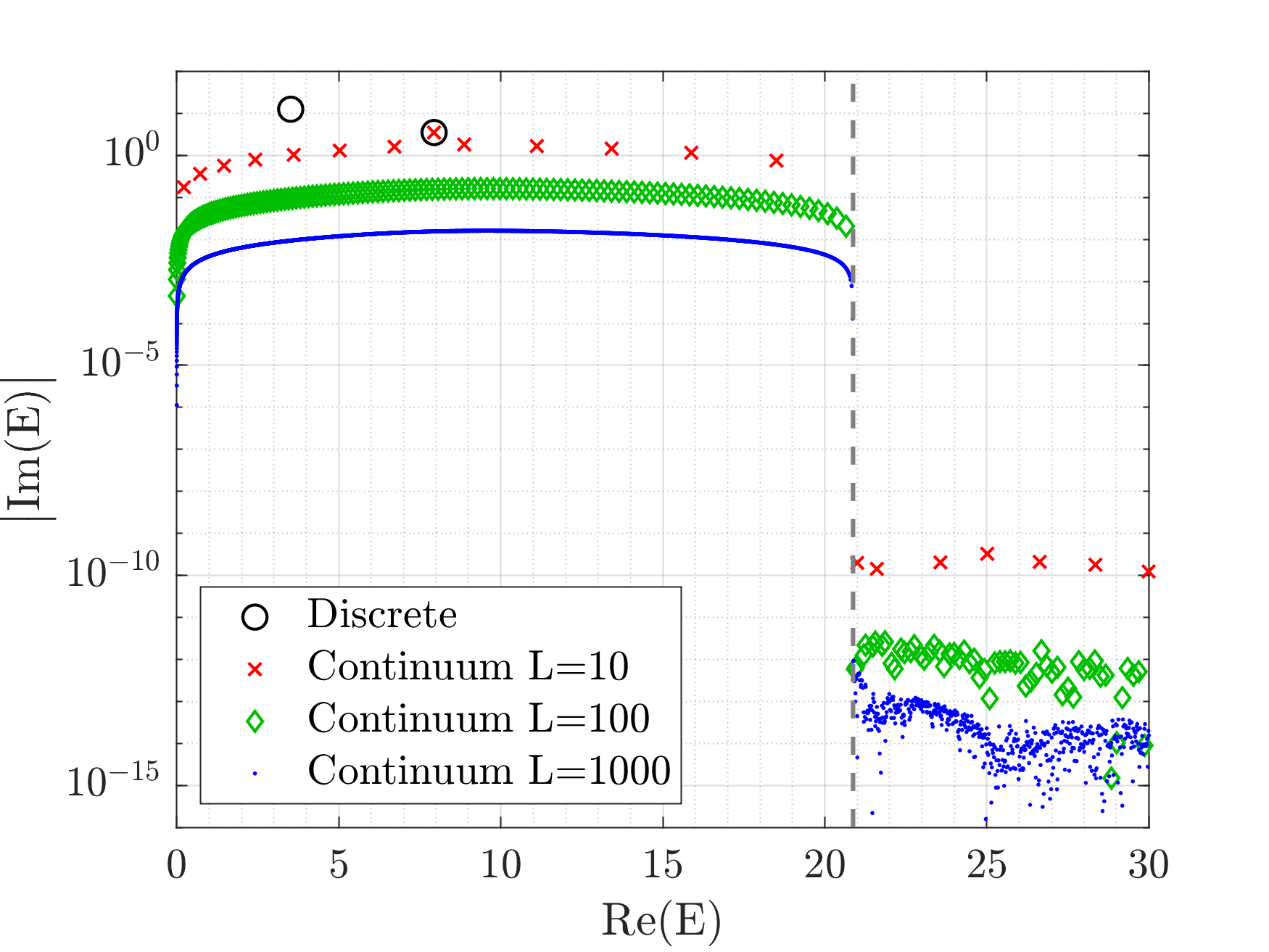}
\end{center}
\caption{[Color online] Logarithmic plot of the energy eigenvalues for the
potential $V_2(x)$ plotted for $L=10$, $100$, and $1000$. Like the eigenvalues
in Fig.~\ref{F3}, the continuum eigenvalues here undergo an abrupt jump at the
dashed line near the critical value close to $21$, where the imaginary parts of
the continuum eigenvalues suddenly drop by about 10 orders of magnitude.
The location of this line is insensitive to the value of $L$.}
\label{F7}
\end{figure}

As with the Scarf-II potential $V_1(x)$, there is a transition in the continuum
part of the spectrum that for this model occurs near $21$. To examine this
transition, we plot the eigenvalues on a logarithmic scale in Fig.~\ref{F7}.
Observe that near $21$ the continuum eigenvalues undergo an abrupt jump in
their imaginary parts of 10 orders of magnitude.

\section{Eigenvalue behavior of $V_3(x)$}
\label{s4}
\begin{figure}[h!]
\begin{center}
\includegraphics[scale=0.50]{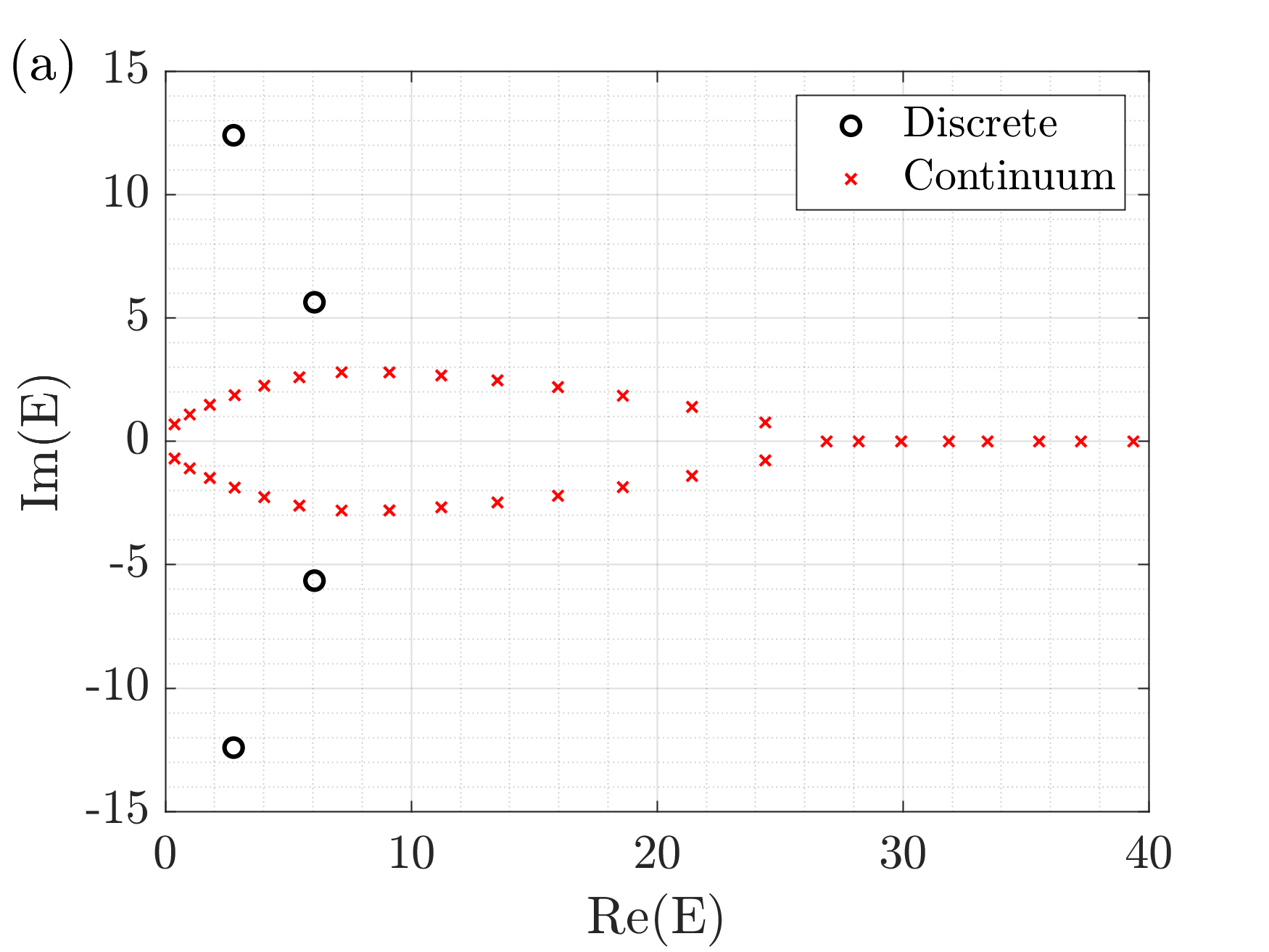}\hspace{-.1cm}
\includegraphics[scale=0.50]{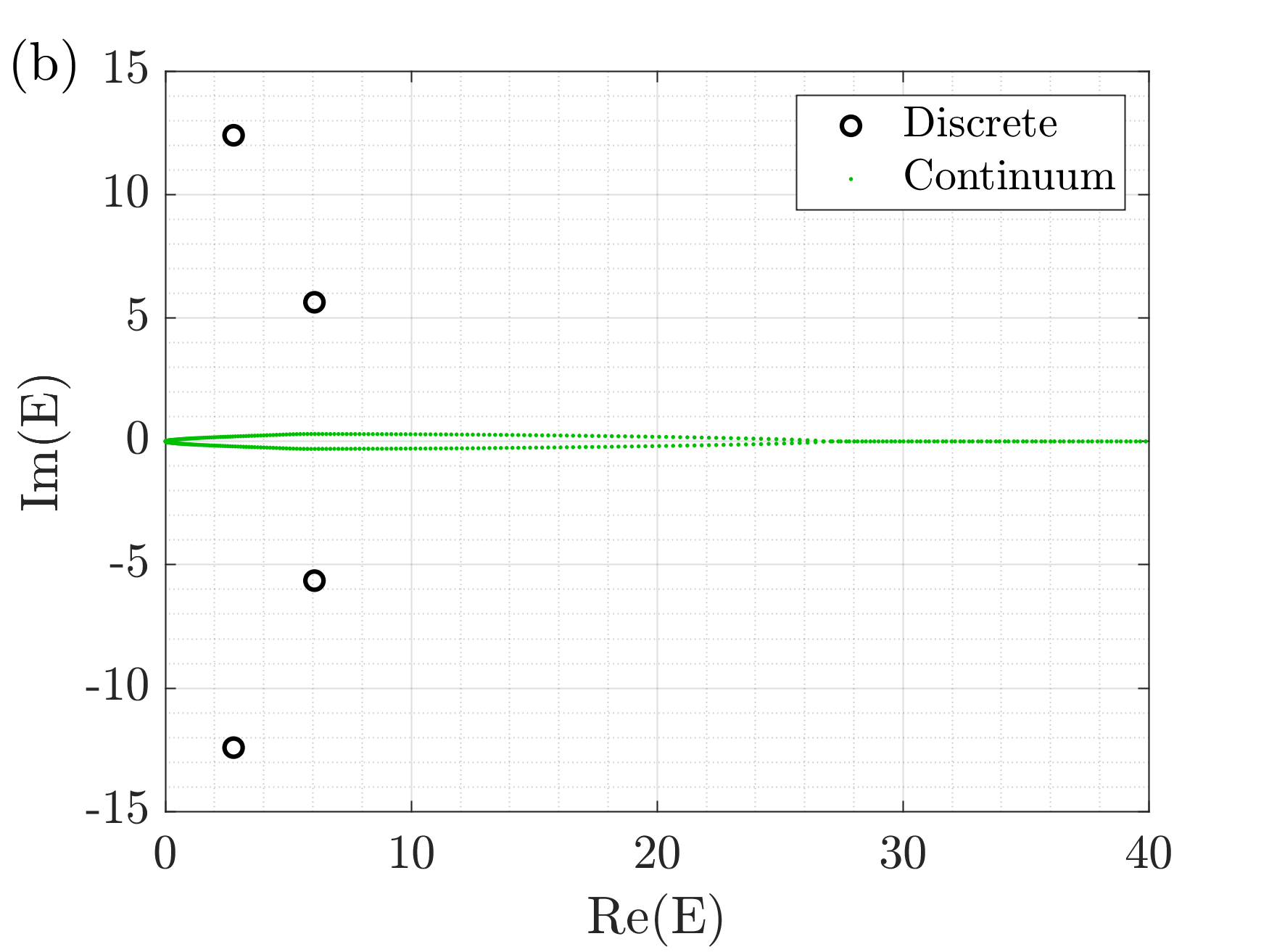}
\end{center}
\caption{[Color online] Energy eigenvalues for $V_3(x)$ with $A_3=30$. In panel
(a) we display the data for $L=10$ and in panel (b) we display the data for
$L=100$. Observe that as $L$ increases, the continuous eigenvalues approach the
real axis. However the discrete complex-conjugate bound-state eigenvalues
(black circles) remain fixed.}
\label{F8}
\end{figure}

The spectral structure associated with the potential $V_3(x)$ is qualitatively
similar to that of $V_2(x)$. We take the strength parameter $A_3=30$ and
plot the eigenvalues for $L=10$ in Fig.~\ref{F8}, panel (a), and for $L=100$
in Fig.~\ref{F8}, panel (b).

\begin{figure}[t!]
\begin{center}
\includegraphics[scale=0.50]{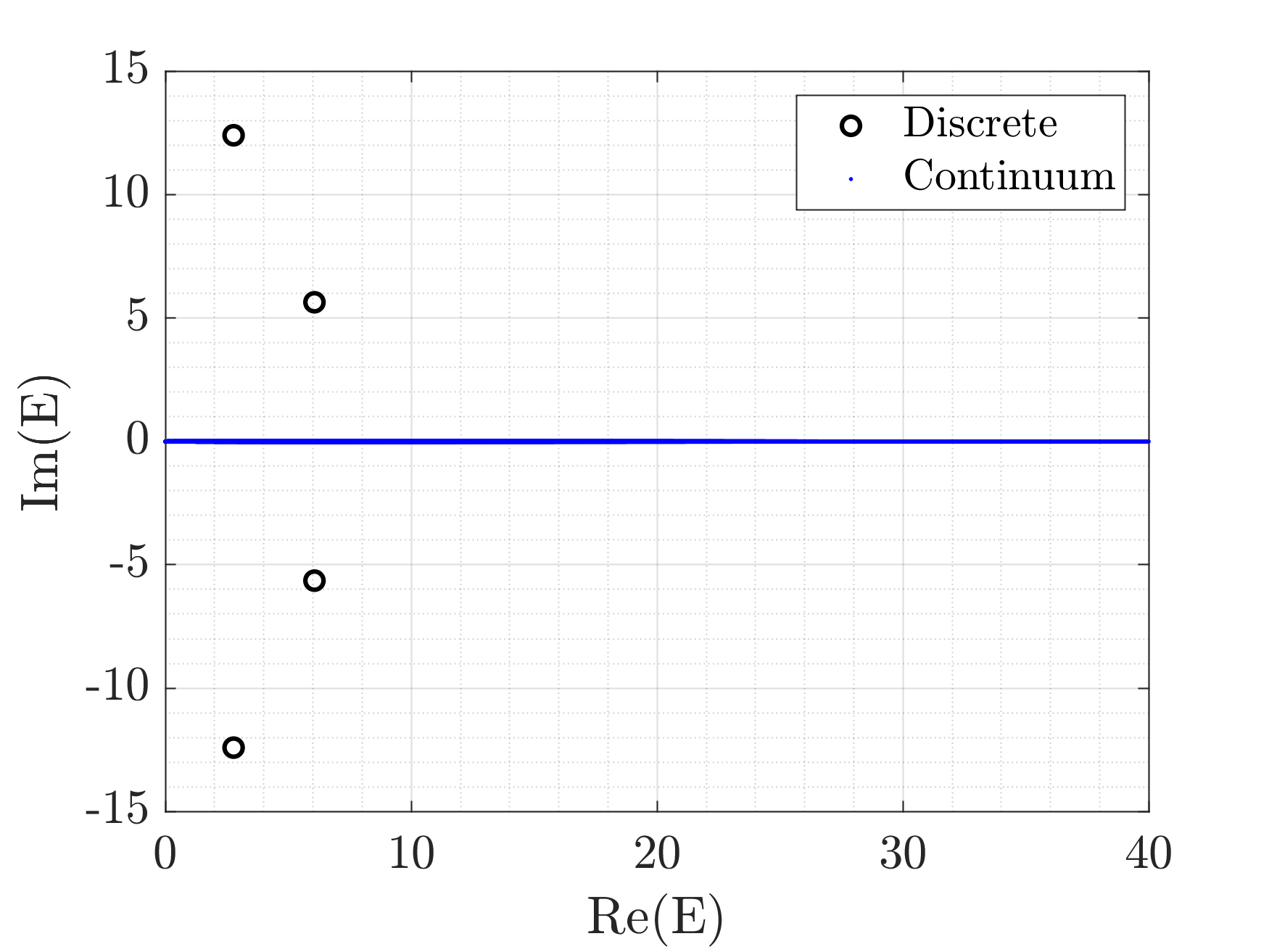}
\end{center}
\caption{[Color online] Energy eigenvalues for $V_3(x)$ with $A_3=30$
calculated at $L=1000$.}
\label{F9}
\end{figure}

\begin{figure}[h!]
\begin{center}
\includegraphics[scale=0.50]{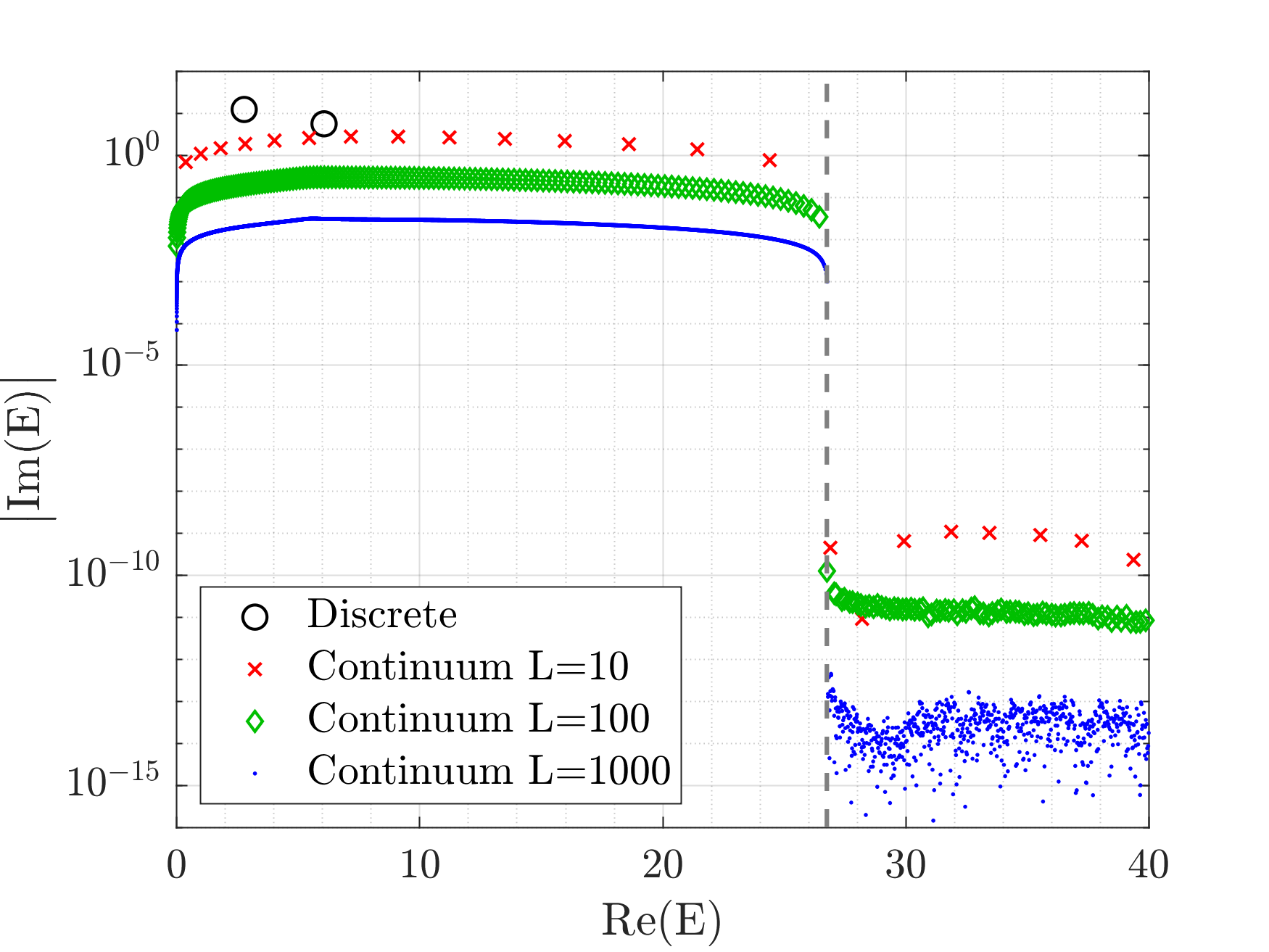}
\end{center}
\caption{[Color online] Logarithmic plot of the data in Figs.~\ref{F8} and
\ref{F9}. As in Figs.~\ref{F3} and \ref{F7} we see once again that there is a
transition point, in this case near 27, at which the imaginary parts of the
continuum eigenvalues suddenly drop by about ten orders of magnitude. The
location of this transition, which is indicated by a dashed line, appears to be
independent of the choice of $L$.}
\label{F10}
\end{figure}

Observe that as $L$ is increased from $10$ to $100$, the discrete bound-state
eigenvalues do not move but the continuum part of the spectrum rapidly
approaches the real axis. In Fig.~\ref{F9} we increase the value of $L$ to 1000.
This higher-accuracy calculation shows that the continuum eigenvalues are
extremely close to the real axis. However, there is still a critical point where
the continuum eigenvalues go from having a small imaginary part to a vanishing
imaginary part. This transition point is near $27$ and the transition is
indicated in Fig.~\ref{F10} by a dashed line. Once again, the logarithmic plot
shows that at the transition the imaginary parts of the continuum eigenvalues
abruptly drop by about ten orders of magnitude.

\section{Eigenvalue behavior of $V_4(x)$}
\label{s5}
\begin{figure}[h!]
\begin{center}
\includegraphics[scale=0.50]{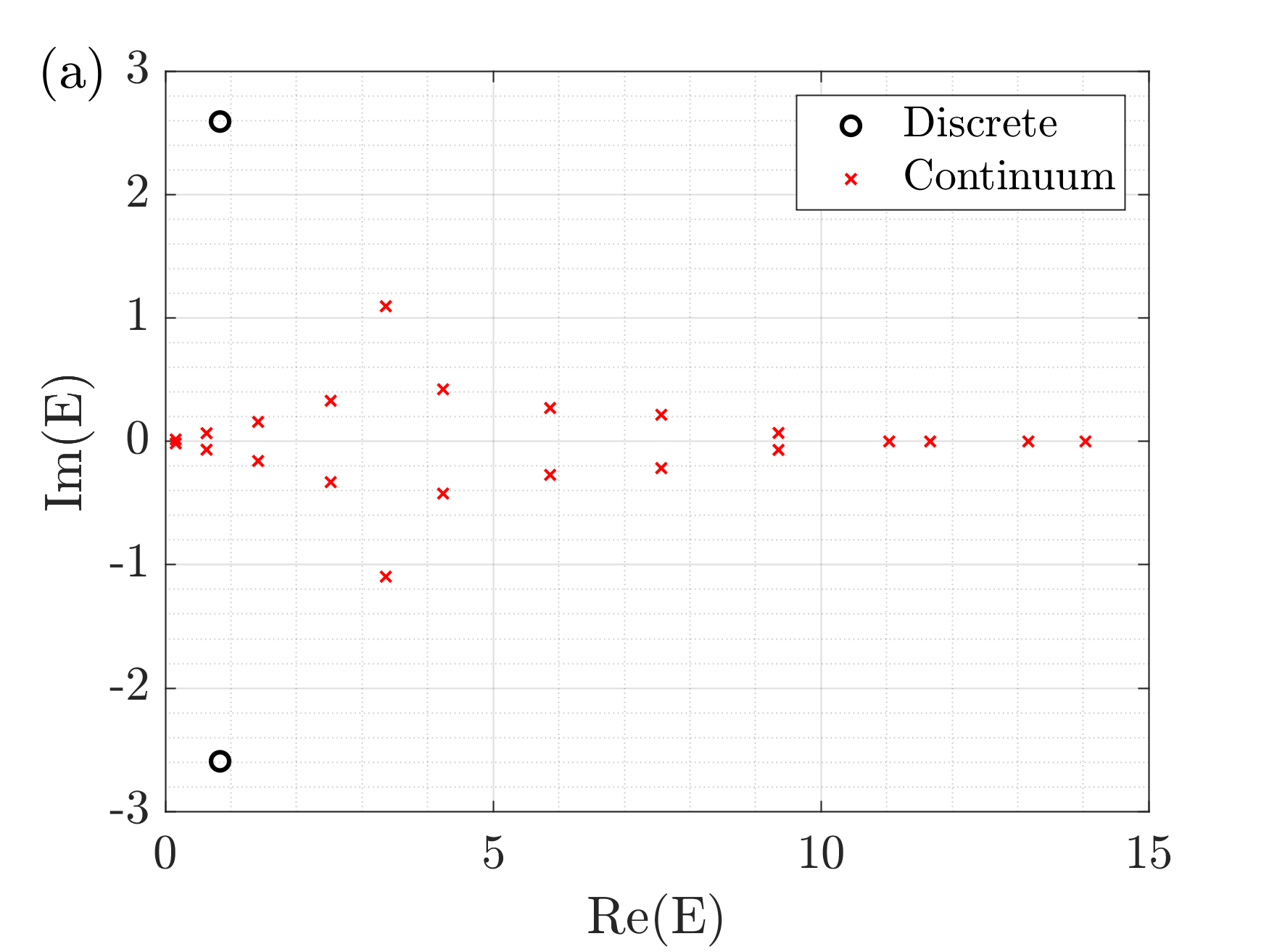}\hspace{-.1cm}
\includegraphics[scale=0.50]{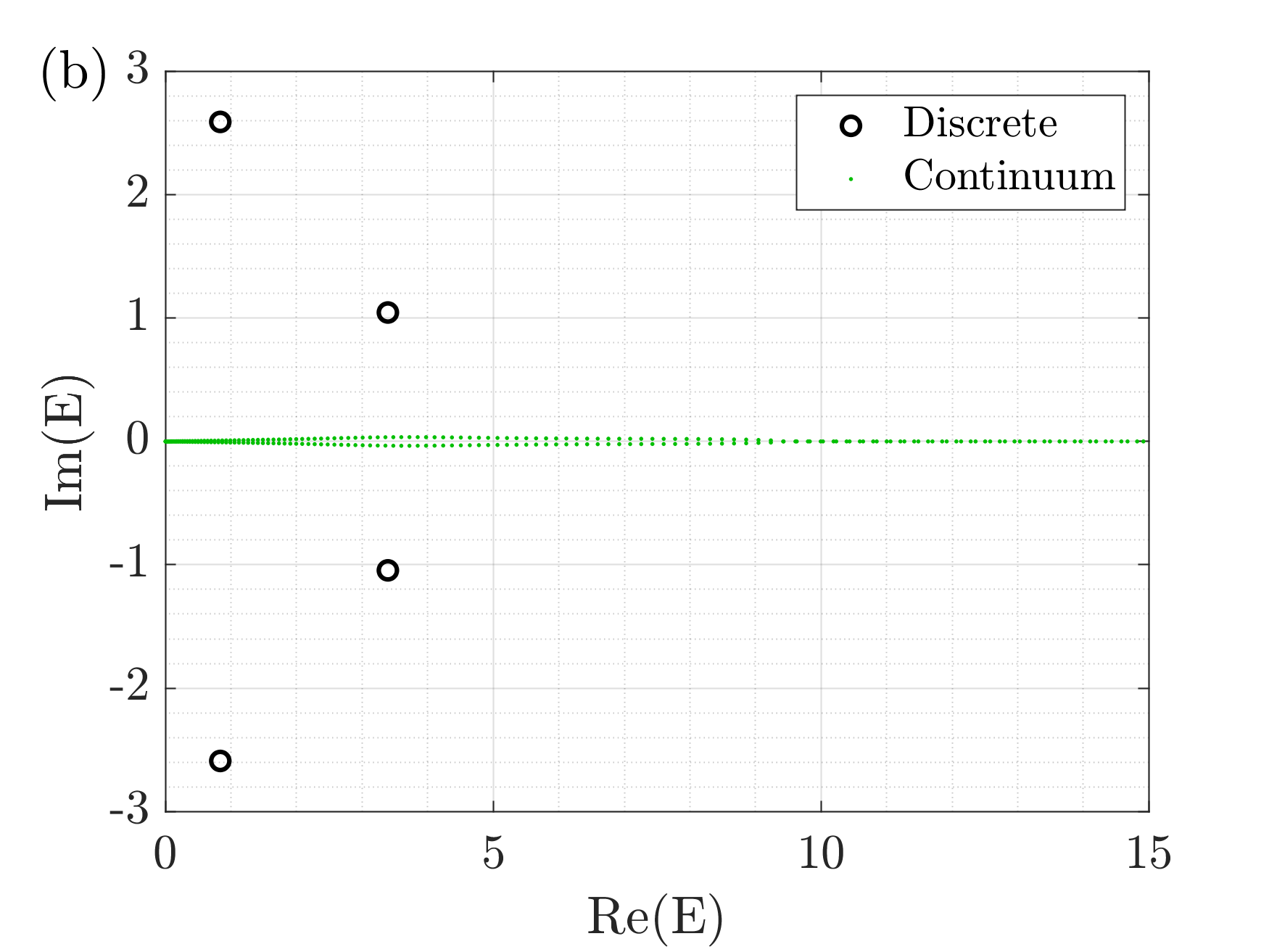}
\end{center}
\caption{[Color online] Energy eigenvalues associated with $V_4(x)$ for $L=10$
in panel (a) and for $L=100$ in panel (b). The strength parameter $A_4=3$. One
pair of bound-state energies (black circles) can be seen in panel (a) but a
new pair becomes visible in Panel (b).}
\label{F11}
\end{figure}

\begin{figure}[h!]
\begin{center}
\includegraphics[scale=0.50]{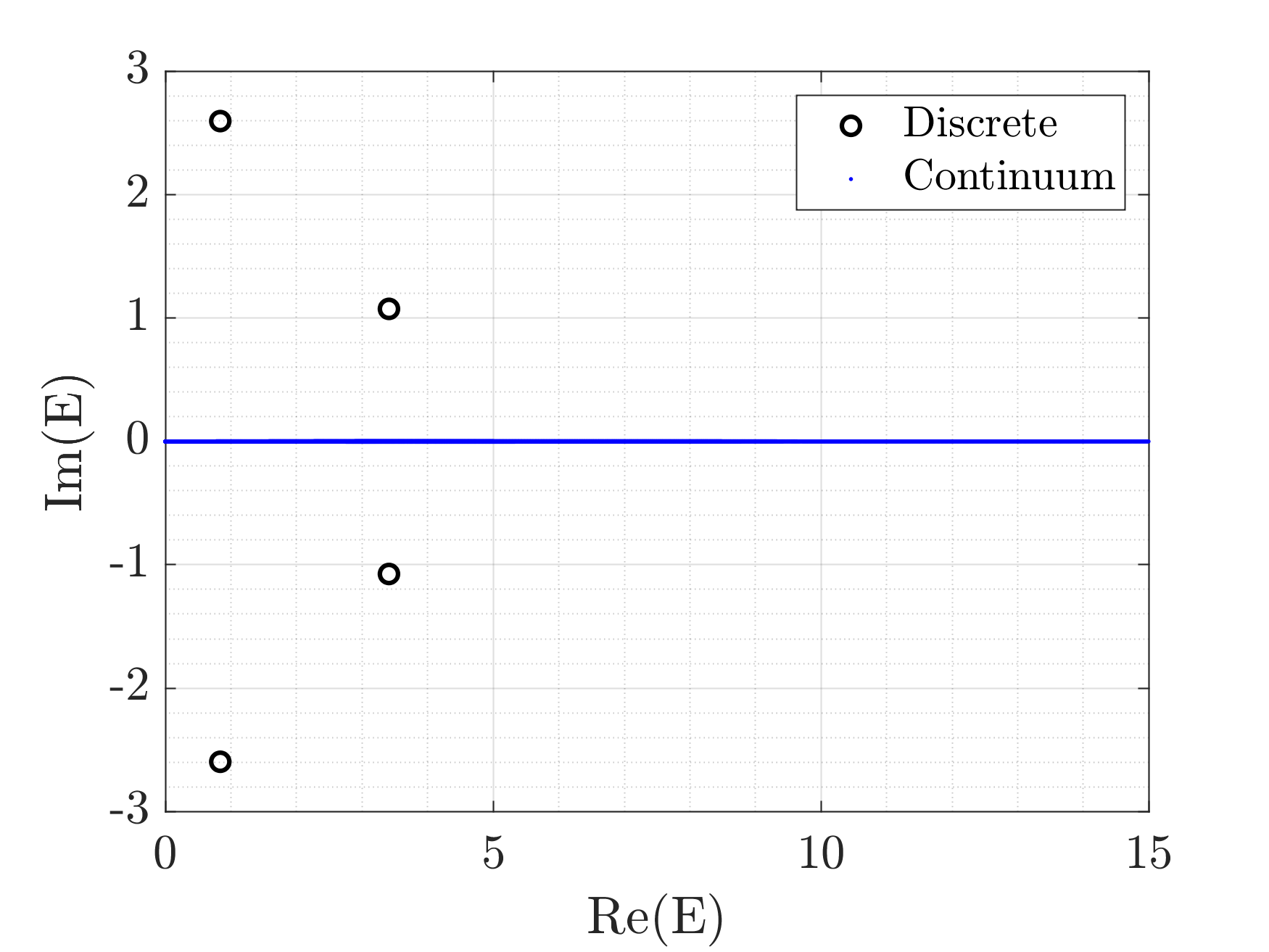}
\end{center}
\caption{[Color online] Energy eigenvalues for $V_4(x)$ with $A_4=3$ for
$L=1000$. Note that the bound-state eigenvalues have not changed from their
position from those in panel (b) of Fig.~\ref{F11}.}
\label{F12}
\end{figure}

The pattern of eigenvalues associated with $V_4(x)$ is similar to that of
$V_1(x)$, $V_2(x)$, and $V_3(x)$. For this potential we take the strength
parameter $A_4=3$ and plot the spectra for $L=10$ and $L=100$ in Fig.~\ref{F11}
and for $L=1000$ in Fig.~\ref{F12}. These figures show no qualitatively new
features.

\begin{figure}[h!]
\begin{center}
\includegraphics[scale=0.50]{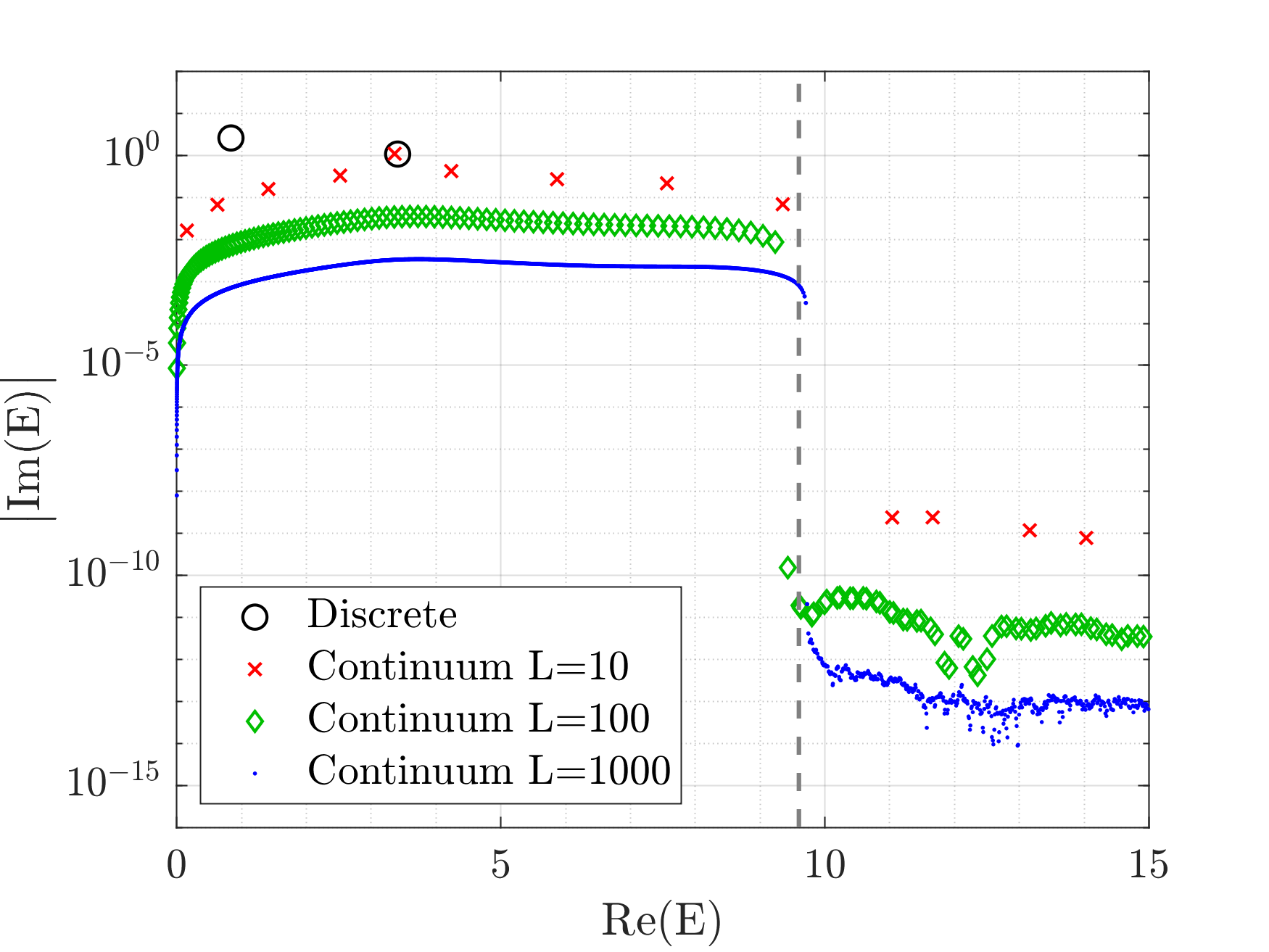}
\end{center}
\caption{[Color online] Logarithmic plot of the eigenvalue data in
Figs.~\ref{F11} and \ref{F12}. The data in this figure, like the data in
Figs.~\ref{F3}, \ref{F7}, and \ref{F10}, indicates that the sharp transition,
which in this case is close to $9.5$ is not sensitive to the value of $L$.}
\label{F13}
\end{figure}

A logarithmic plot of the eigenvalue data in Figs.~\ref{F11} and \ref{F12} is 
shown in Fig.~\ref{F13}. Once again we see a transition, in this case near
$9.5$, at which the continuum eigenvalues abruptly drop in magnitude by about
ten orders of magnitude. The location of the transition is again insensitive to
the value of $L$.

\section{Eigenvalue behavior of $V_5(x)$}
\label{s6}
\begin{figure}[h!]
\begin{center}
\includegraphics[scale=0.50]{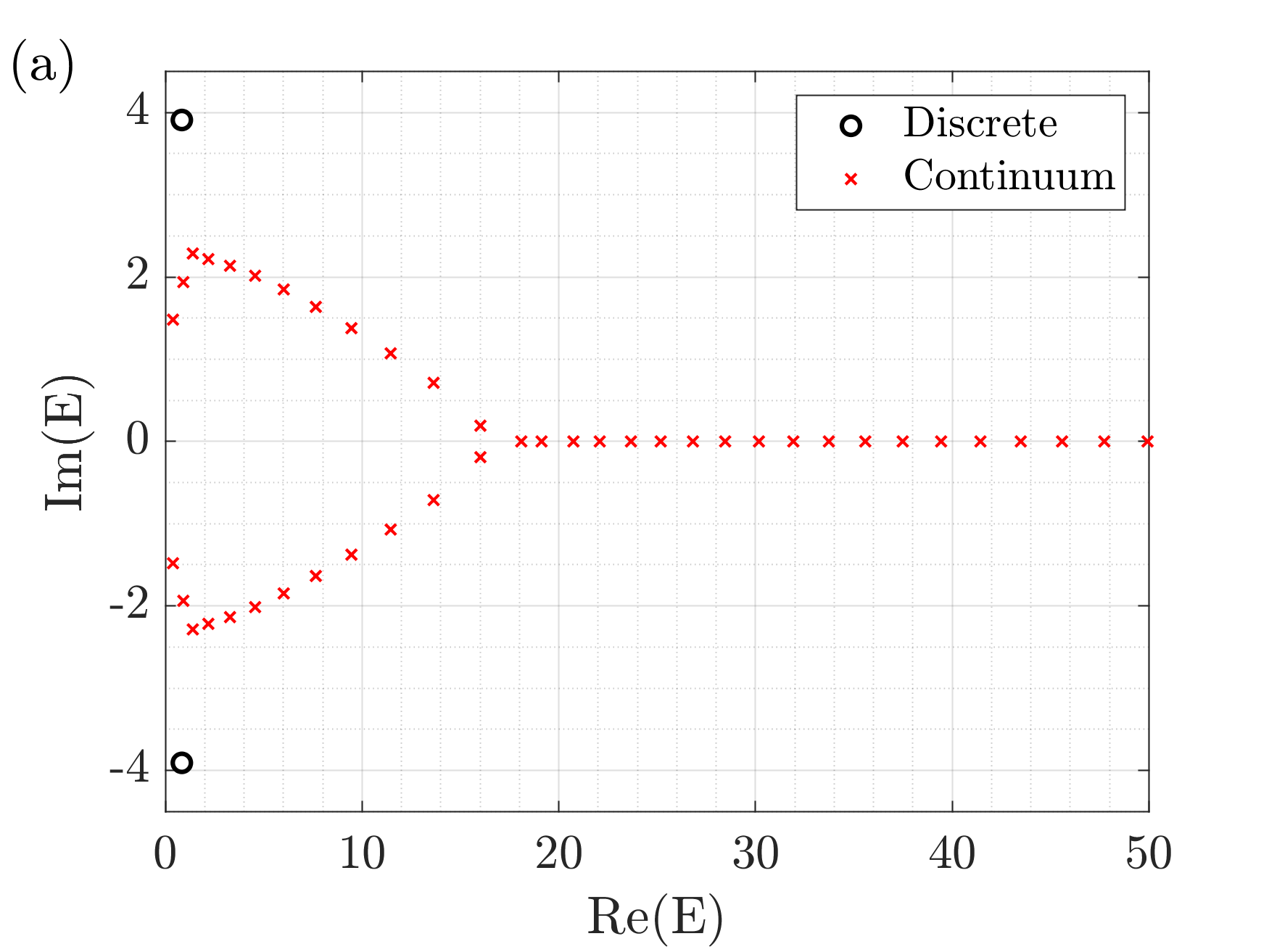}\hspace{-.1cm}
\includegraphics[scale=0.50]{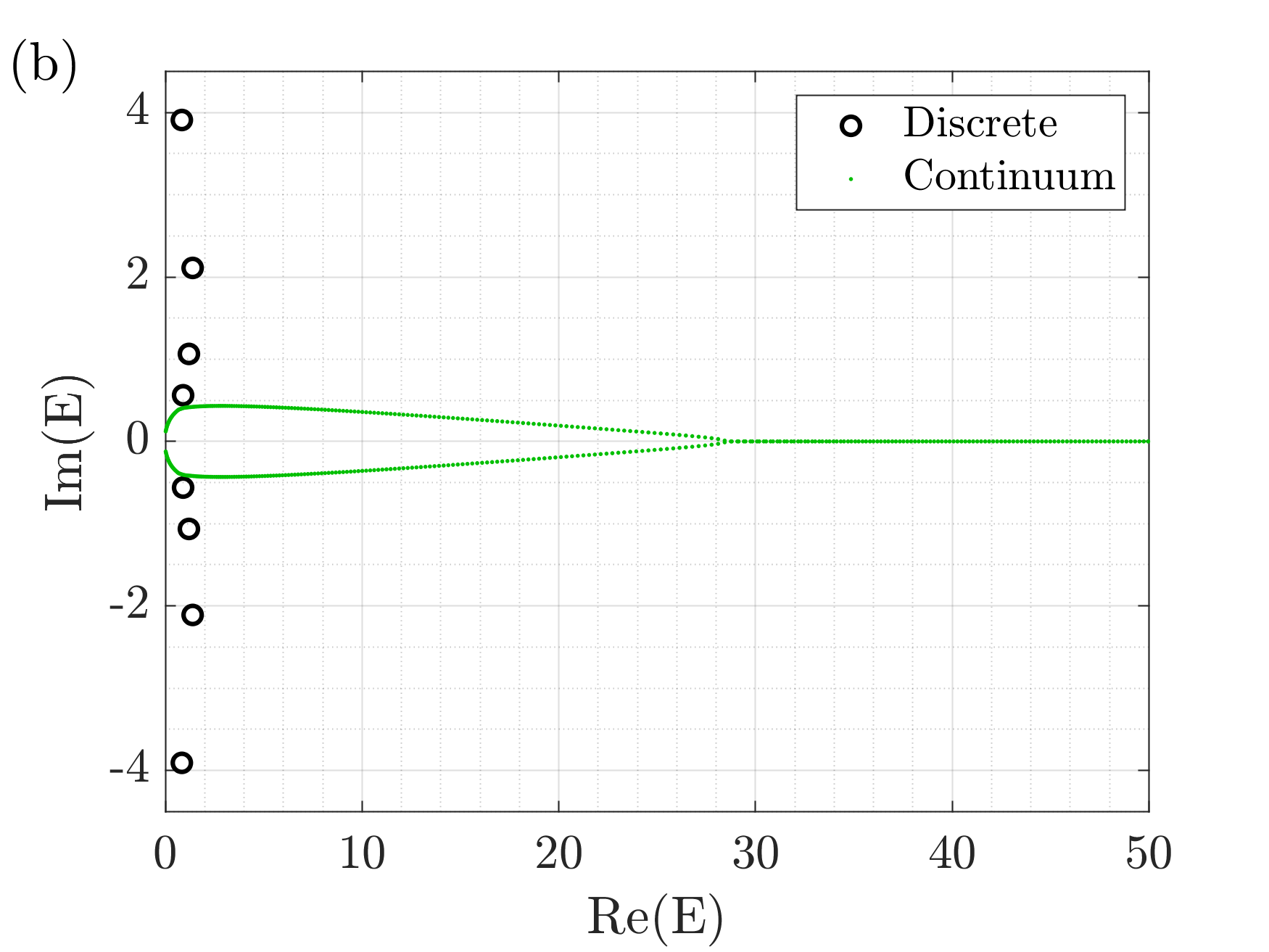}
\end{center}
\caption{[Color online] Energy eigenvalues associated with $V_5(x)$ with
strength parameter $A_5=10$. In panel (a) we take $L=10$ and we observe one
complex-conjugate pair of bound-state eigenvalues; in panel (b) we take $L=100$
and observe three new pairs of complex bound-state eigenvalues. Unlike the
results for short-range potentials, this figure shows that the sequence of
bound-state eigenvalues is turning around and heading back towards the
origin. Also, the transition points in the continuum part of the spectrum
are not fixed but are moving up the real axis from about $16$ in panel (a)
to about $28$ in panel (b).}
\label{F14}
\end{figure}

The most interesting and surprising results that we have obtained concern the 
eigenspectrum associated with $V_5(x)$. For this potential we take the strength
parameter $A_5=10$. Because this is a long-range potential, it is not easy to
obtain accurate and trustworthy numerical results, and we have had to do the
$L=1000$ calculation in {\it quadruple} precision (for all other results in
this paper double precision is sufficient). In Fig.~\ref{F14} we plot the
eigenvalues for $L=10$ in panel (a) and $L=100$ in panel (b). There is one pair
of bound-state eigenvalues in panel (a). When we increase the size of the
interval, we see in panel (b) that the continuum spectrum has dropped much
closer to the real axis and has uncovered three new pairs of bound-state
eigenvalues.

Figure~\ref{F14} reveals two new effects that we have not observed in our
studies of short-range potentials. First, the sequence of bound-state
eigenvalues has {\it turned around} and is heading backward towards the origin.
In Figs.~\ref{F1}, \ref{F5}, \ref{F8}, and \ref{F11} the real parts of the
eigenvalues are increasing, not decreasing. Second, the transition in the
continuum part of the spectrum at which the eigenvalues become real is no
longer a fixed point on the real axis; rather, the transition point is moving
up the real axis as $L$ increases. In panel (a) the transition is near
$16$ but in panel (b) it is near $28$.

\begin{figure}[h!]
\begin{center}
\includegraphics[scale=0.50]{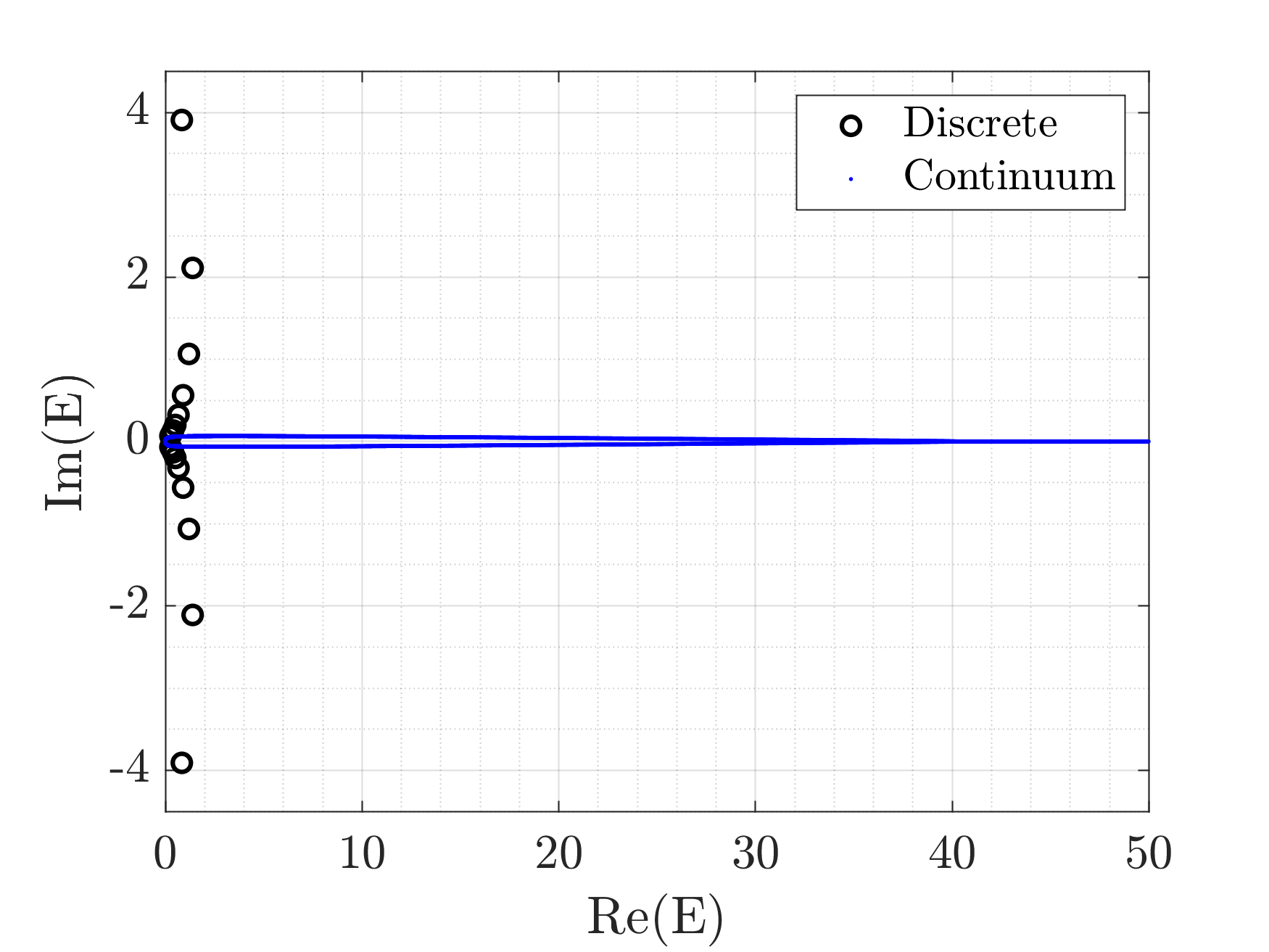}
\end{center}
\caption{[Color online] Energy eigenvalues associated with $V_5(x)$ with
strength parameter $A_5=10$. In this figure we have have increased $L$ from 
$100$ in Fig.~\ref{F14} (right panel) to $1000$ and there are now nine complex
pairs of bound-state energies (not easy to distinguish). The numerical
calculations needed to produce this figure required quadruple precision.}
\label{F15}
\end{figure}

\begin{figure}[h!]
\begin{center}
\includegraphics[scale=0.50]{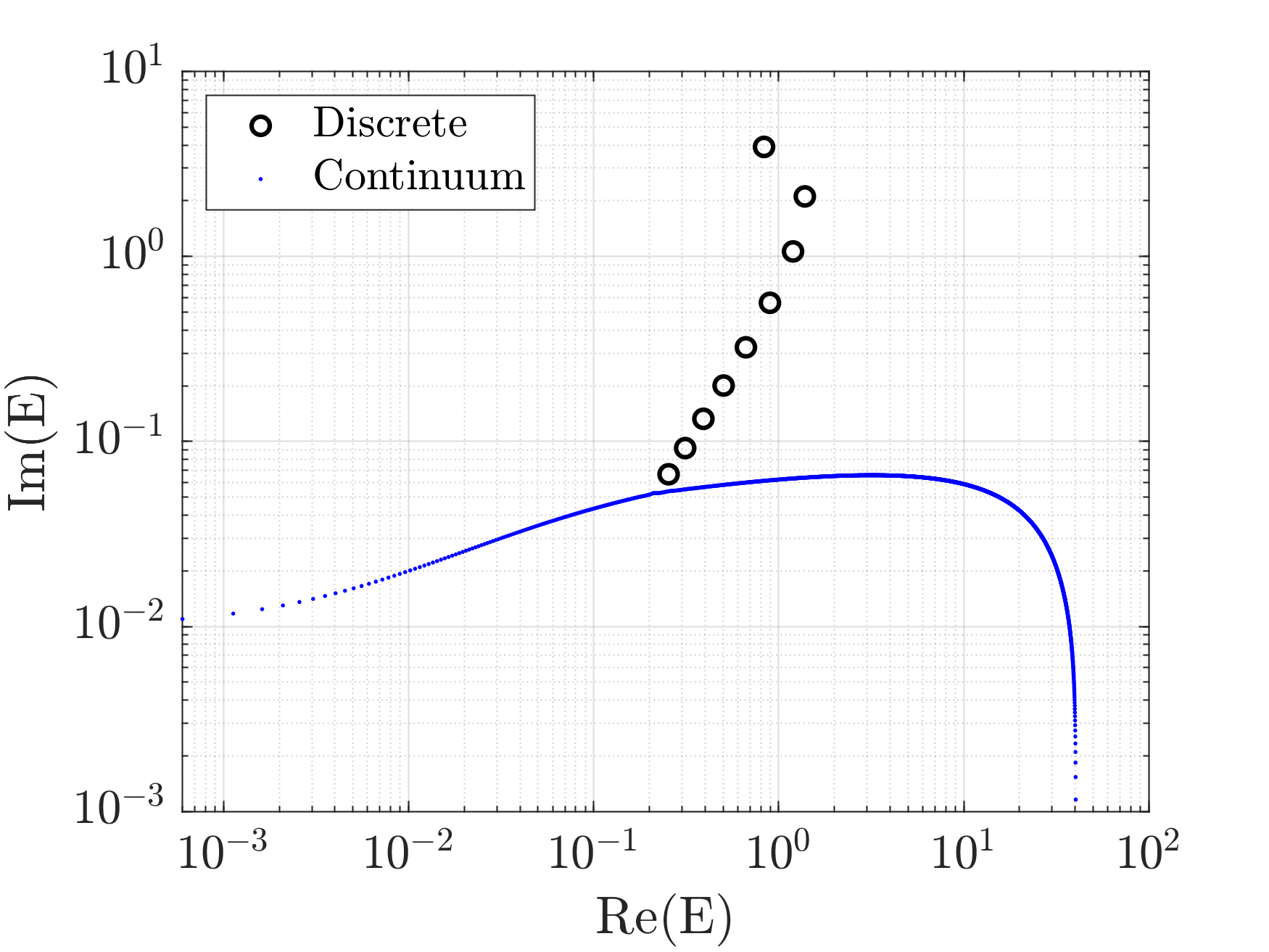}
\end{center}
\caption{[Color online] Plot of the eigenvalue data in Fig.~\ref{F15} on a
log-log graph. In this plot we can now easily see nine bound-state eigenvalues.}
\label{F16}
\end{figure}

If we increase $L$ to $1000$, Fig.~\ref{F15} shows that there are now {\it
nine} complex-conjugate pairs of bound-state eigenvalues (which are not easy
to see clearly). This sequence of bound-state eigenvalues is tending towards the
origin. To observe the bound-state eigenvalues more clearly we have replotted in
Fig.~\ref{F16} the data in Fig.~\ref{F15} on a log-log plot. We can see on
this plot that the sequence of bound-state eigenvalues is becoming linear. 

\begin{figure}[h!]
\begin{center}
\includegraphics[scale=0.50]{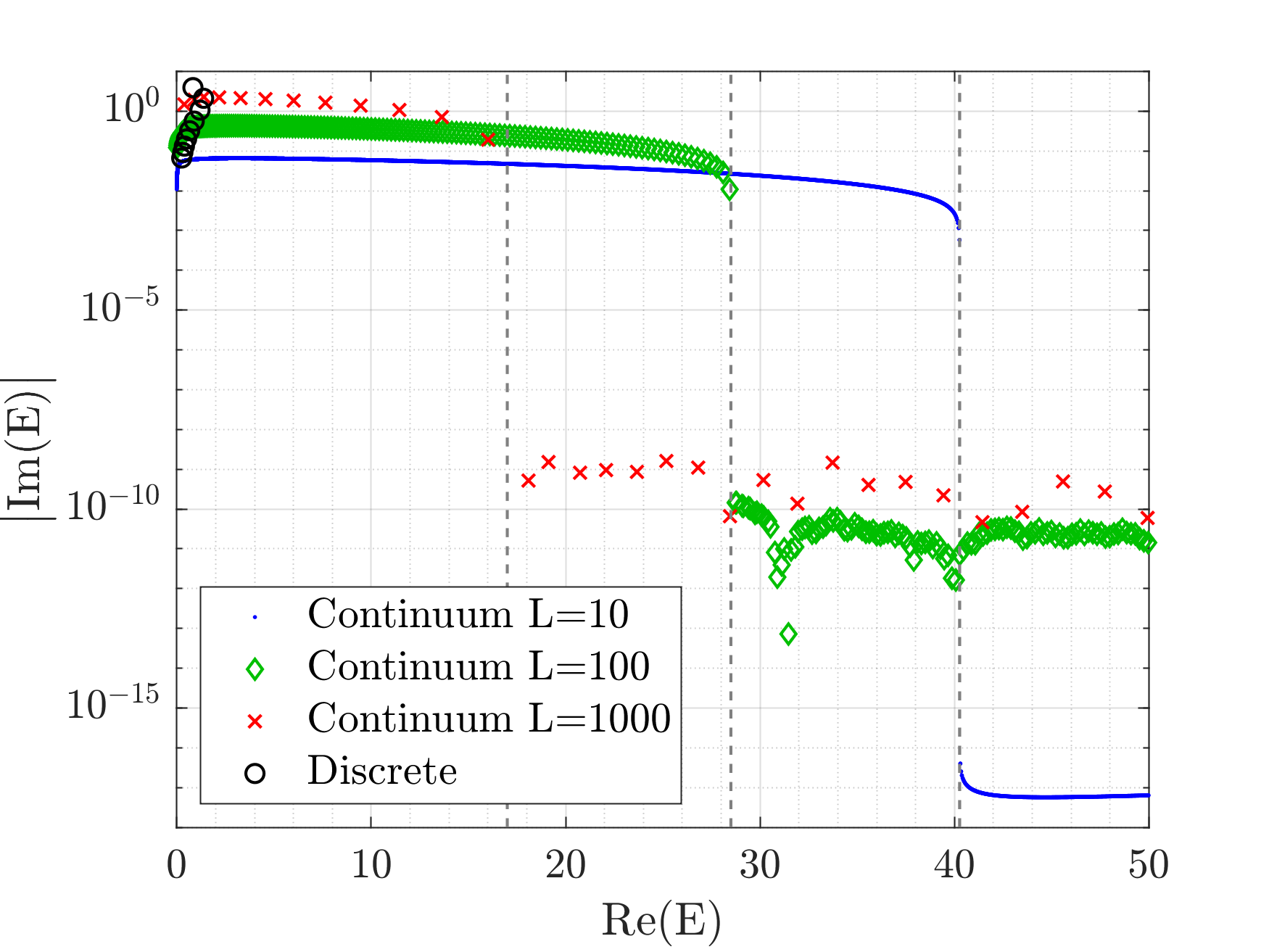}
\end{center}
\caption{[Color online] Logarithmic plot of the eigenvalue data in
Figs.~\ref{F14} and \ref{F15}. Observe that the transition point in the
continuum part of the spectrum is moving up the real axis as $L$ increases
from $L=10$ to $L=100$ to $L=1000$.}
\label{F17}
\end{figure}

To observe the transition points in the continuum part of the eigenspectrum,
we have plotted the data in Figs.~\ref{F14} and \ref{F15} on a logarithmic graph
in Fig.~\ref{F17}. Note that the transition point in the continuum eigenvalues
from complex to real is moving up the real axis; it is no longer fixed as it was
for finite-range potentials. Observe that at the transition near $40$ there is a
drop of nearly $20$ (and not $10$) orders of magnitude. To see this effect
requires that we use quadruple and not double precision in our numerical
calculations.

The most interesting aspect of the long-range potential $V_5(x)$ is that it
appears to confine an infinite number of bound states and the complex
bound-state energies appear to be approaching $0$. To verify this we use
Richardson extrapolation \cite{R14} to study the behavior of the sequence of
bound-state energies.

Richardson extrapolation enables one to find the limit of the sequence $\{a_k\}$
as $k\to\infty$ if the limit is a finite number. Given such a sequence we can
calculate more and more accurate Richardson extrapolants, which converge faster
to the limiting value. The formulas for the first five Richardson extrapolants
are given by
\begin{eqnarray}
R_k^{(1)}&=&(k+1)a_{k+1}-ka_k,\nonumber\\
R_k^{(2)}&=&\left[(k+2)^2a_{k+2}-2(k+1)^2a_{k+1}+k^2a_k\right]/2!,\nonumber\\
R_k^{(3)}&=&\left[(k+3)^3a_{k+3}-3(k+2)^3a_{k+2}+3(k+1)^3a_{k+1}-k^3a_k\right]/
3!,\nonumber\\
R_k^{(4)}&=&\left[(k+4)^4a_{k+4}-4(k+3)^4a_{k+3}+6(k+2)^4a_{k+2}\right.
\nonumber\\
&&\left.\qquad\qquad-4(k+1)^4a_{k+1}+k^4a_k\right]/4!,\nonumber\\
R_k^{(5)}&=&\left[(k+5)^5a_{k+5}-5(k+4)^5a_{k+4}+10(k+3)^5a_{k+3}\right.
\nonumber\\
&&\left.\qquad\qquad-10(k+2)^5 a_{k+2}+5(k+1)^5a_{k+1}+k^5a_k\right]/5!.
\label{e2}
\end{eqnarray}

From our numerical analysis, we have determined that the $k$th bound-state
energy $E_k$ has the asymptotic form 
\begin{equation}
E_k\sim\frac{\alpha}{k^2}\pm i\frac{\beta}{k^3}\qquad(k\gg1),
\label{e3}
\end{equation}
where $\alpha$ and $\beta$ are real numbers. This shows that the bound-state
eigenvalues associated with $V_5(x)$ share many of the quantitative features of
the Balmer series for the hydrogen atom. As indicated in the tables \ref{t1} and
\ref{t2}, we have determined that the numerical value of $\alpha$ is about $25$
and the numerical value of $\beta$ is about $61$. To obtain these results we
have multiplied the real part of $E_k$ by $k^2$ and the imaginary part of
$E_k$ by $k^3$ and computed the first five Richardson extrapolations.

\begin{table}[h!]
\begin{center}
\begin{tabular}{|c|c|c|c|c|c|c|c|}
\hline
$k$ & ${\rm Re}\,E_k$ & $k^2\,{\rm Re}\,E_k$ & $R_k^{(1)}$ & $R_k^{(2)}$ &
$R_k^{(3)}$ & $R_k^{(4)}$ & $R_k^{(5)}$ \\
\hline
1& 0.83298288 & 0.83298 & 10.2355 & 26.6628 & 29.8431 & 23.8084 & 24.2927\\
2& 1.38356468 & 5.53426 & 21.1871 & 29.0481 & 25.0154 & 24.2120 & 25.5106\\
3& 1.19465086 & 10.7519 & 25.1176 & 26.6284 & 24.4798 & 25.1396 & 25.2280\\
4& 0.89645517 & 14.3433 & 25.7219 & 25.5541 & 24.8568 & 25.1949 & 25.0599\\
5& 0.66476032 & 16.6190 & 25.6660 & 25.2553 & 25.0258 & 25.1199 &    --- \\
6& 0.50352322 & 18.1268 & 25.5486 & 25.1692 & 25.0676 &    ---  &    --- \\
7& 0.39157331 & 19.1871 & 25.4538 & 25.1354 &    ---  &    ---  &    --- \\
8& 0.31203794 & 19.9704 & 25.3830 &    ---  &    ---  &    ---  &    --- \\
9& 0.25397318 & 20.5718 &    ---  &    ---  &    ---  &    ---  &    --- \\
\hline
\end{tabular}
\end{center}
\caption{\label{t1} Real parts of the first nine eigenvalues $E_k$ $(1\leq k\leq
9)$ associated with $V_5(x)$ and the first five Richardson extrapolants
constructed from the sequence $\{k^2\,{\rm Re}\,E_k\}$. Evidently, the real
parts of the eigenvalues vanish like $\alpha k^{-2}$, where $\alpha$ is roughly
$25$.}
\end{table}

\begin{table}[h!]
\begin{center}
\begin{tabular}{|c|c|c|c|c|c|c|c|}
\hline
$k$ & ${\rm Im}\,E_k$ & $k^3\,{\rm Im}\,E_k$ & $R_k^{(1)}$ & $R_k^{(2)}$ &
$R_k^{(3)}$ & $R_k^{(4)}$ & $R_k^{(5)}$ \\
\hline
1& 3.90859038 & 3.90859 & 29.8484 & 63.7004 & 62.5280 & 53.5920 & 61.0166\\
2& 2.10981263 & 16.8785 & 52.4164 & 62.8211 & 55.3792 & 59.7791 & 62.5903\\
3& 1.06386938 & 28.7245 & 57.6188 & 58.3560 & 58.3125 & 61.7871 & 61.3005\\
4& 0.56168819 & 35.9480 & 57.9136 & 58.3342 & 60.2980 & 61.4830 & 60.9267\\
5& 0.32272930 & 40.3412 & 58.0538 & 59.1758 & 60.8905 & 61.1739 &    --- \\
6& 0.20043182 & 43.2933 & 58.3744 & 59.8188 & 61.0165 &    ---  &    --- \\
7& 0.13250064 & 45.4477 & 58.7355 & 60.2180 &    ---  &    ---  &    --- \\
8& 0.09200917 & 47.1087 & 59.0650 &    ---  &    ---  &    ---  &    --- \\
9& 0.06644330 & 48.4372 &    ---  &    ---  &    ---  &    ---  &    --- \\
\hline
\end{tabular}
\end{center}
\caption{\label{t2} Imaginary parts of the first nine eigenvalues $E_k$ $(1\leq
k\leq9)$ associated with $V_5(x)$ and the first five Richardson extrapolants
constructed from the sequence $\{k^3\,{\rm Im}\,E_k\}$. The imaginary parts of
the eigenvalues vanish like $\beta k^{-3}$, where $\beta$ is roughly $61$.}
\end{table}

\section{Conclusions}
\label{s7}

In this paper we have studied numerically the five $\cPT$-symmetric potentials
in (\ref{e1}) that have continuous spectra. Each of these potentials is pure
imaginary and vanishes as $|x| \to\infty$. The interesting feature of these
potentials is that even though they vanish at $\pm\infty$, they still confine
bound states. Of course, an imaginary potential can confine bound states. For
example, the $ix^3$ potential has an infinite number of bound states \cite{R1}.
However, this potential becomes {\it stronger} as $|x|\to\infty$. We emphasize
that the five potentials that we have studied here {\it decay} and become {\it
weaker} as $|x|\to\infty$. Even more remarkable is the fact that the potential
$V_5(x)$, which decays very slowly as $|x|=\infty$, binds an {\it infinite}
number of bound states and that the sequence of bound-state energies
asymptotically approaches the Balmer series for the hydrogen atom.

\vspace{0.5cm}
\footnotesize
\noindent
CMB thanks the Alexander von Humboldt Foundation for partial financial support.

\end{document}